\documentclass[aps,pra,reprint,superscriptaddress]{revtex4-2}

\usepackage{lmodern}
\usepackage[T1]{fontenc}
\usepackage[english,activeacute]{babel}
\usepackage{mathtools}
\usepackage{color} 

\usepackage{amssymb} 
\usepackage{dsfont} 

\usepackage{pbsi}

\usepackage[colorlinks,linkcolor=blue,citecolor=blue,urlcolor=blue]{hyperref}

\addto\extrasenglish{}
\addto\extrasenglish{}

\begin{document}


\newcommand{\dd}{\mbox{d}}
\newcommand{\ii}{\mbox{i}}
\newcommand{\ddd}{\mbox{\scriptsize{d}}}
\newcommand{\iii}{\mbox{\scriptsize{i}}}

\newcommand{\ee}{{\rm{e}}}

\newcommand{\QE}{\mbox{\tiny{QE}}}
\newcommand{\LL}{\mbox{\tiny{L}}}
\newcommand{\C}{\mbox{\tiny{C}}}
\newcommand{\CC}{\mbox{\tiny{C}}}
\newcommand{\EE}{\mbox{\tiny{E}}}

\newcommand{\LP}{\mbox{\tiny{LP}}}
\newcommand{\UP}{\mbox{\tiny{UP}}}

\newcommand{\JC}{\mbox{\tiny{JC}}}

\newcommand{\g}{\mbox{g}}
\newcommand{\e}{\mbox{e}}
\newcommand{\gmini}{\mbox{\scriptsize{g}}}
\newcommand{\emini}{\mbox{\scriptsize{e}}}
\newcommand{\eg}{\mbox{\scriptsize{eg}}}

\newcommand{\eff}{\mbox{\scriptsize{eff}}}

\newcommand{\intt}{\mbox{\scriptsize{int}}}

\newcommand{\Tr}{\mbox{Tr}}

\newcommand{\HH}{\mbox{\tiny{H}}}
\newcommand{\VV}{\mbox{\tiny{V}}}

\newcommand{\II}{\mbox{\tiny{I}}}

\newcommand{\rr}{\mbox{\scriptsize{rad}}} 
\newcommand{\nr}{\mbox{\scriptsize{nrad}}} 
\newcommand{\deph}{\mbox{\scriptsize{deph}}} 

\newcommand{\dtQE}{\tilde{\Delta}_{\QE}} %
\newcommand{\dtC}{\tilde{\Delta}_{\C}} %

\newcommand{\Ed}{\boldsymbol{E}_{\rm D}}

\newcommand{\funAAg}{g_{\boldsymbol A ^2} }%
\newcommand{\funAAf}{f_{\boldsymbol A ^2}}%

\newcommand{\funApg}{g_{ \boldsymbol A \cdot \boldsymbol p}}
\newcommand{\funApf}{f_{ \boldsymbol A \cdot \boldsymbol p}}
\newcommand{\funtildeApf}{\tilde{f}_{{\boldsymbol A} \cdot {\boldsymbol p}}}


\newcommand{\bp}{{\boldsymbol{p}}}
\newcommand{\bA}{{\boldsymbol{A}}}
\newcommand{\bR}{{\boldsymbol{R}}}

\newcommand{\bmu}{{\boldsymbol{\mu}}}

\newcommand{\hatmum}{\hat {{\mu} }_{\rm m}}
\newcommand{\hatbr}{\hat{\boldsymbol r}}
\newcommand{\hatbp}{\hat{\boldsymbol p}}
\newcommand{\hatbmu}{\hat{\boldsymbol \mu}}

\newcommand{\hatbA}{\hat{\boldsymbol A}}

\newcommand{\bE} {{\boldsymbol{E}}}
\newcommand{\bB} {{\boldsymbol{B}}}

\newcommand{\bu}{\boldsymbol{u}}
\newcommand{\hatbu}{\hat{\boldsymbol{u}}}

\newcommand{\bx}{{\boldsymbol{x}}}
\newcommand{\br}{{\boldsymbol{r}}}
\newcommand{\bk}{{\boldsymbol{k}}}
\newcommand{\bxi}{{\boldsymbol{\xi}}}

\newcommand{\bj}{{\boldsymbol{j}}}
\newcommand{\bv}{{\boldsymbol{v}}}

\newcommand{\brho}{\boldsymbol{\rho}}

\newcommand{\hatakl}{\hat{a}_{\bk,\lambda}}
\newcommand{\hatakldag}{\hat{a}_{\bk,\lambda}^\dagger}

\newcommand{\beps}{\boldsymbol{\epsilon}}
\newcommand{\epskl}{\varepsilon_{\bk \lambda}}
\newcommand{\epsl}{\varepsilon_{\lambda}}
\newcommand{\bepskl}{\boldsymbol{\varepsilon}_{\bk \lambda}}
\newcommand{\bepsl}{\boldsymbol{\varepsilon}_{\lambda}}
	
\newcommand{\bukl}{\boldsymbol{u}_{\bk,\lambda}}
\newcommand{\buklast}{\boldsymbol{u}^\ast_{\bk,\lambda}}

\newcommand{\bul}{\boldsymbol{u}_{\lambda}}
\newcommand{\bulast}{\boldsymbol{u}^\ast_{\lambda}}

\definecolor{internationalorange}{rgb}{1.0, 0.31, 0.0}

\newcommand{\pr}[1]{\textcolor{blue}{#1}}
\newcommand{\rsb}[1]{\textcolor{internationalorange}{#1}}

\makeatletter
\let\oldtheequation\theequation
\renewcommand\tagform@[1]{\maketag@@@{\ignorespaces#1\unskip\@@italiccorr}}
\renewcommand\theequation{(\oldtheequation)}
\makeatother
\renewcommand{\equationautorefname}{Eq.}
\renewcommand{\figureautorefname}{Fig.}
	
\title{Can we observe non-perturbative vacuum shifts in cavity QED?}

\author{Roc\'io S\'aez-Bl\'azquez}
\affiliation{Vienna Center for Quantum Science and Technology,
	Atominstitut, TU Wien, 1020 Vienna, Austria}

\author{Daniele de Bernardis}
\affiliation{INO-CNR BEC Center and Dipartimento di Fisica, Universit\`a di Trento, I-38123 Povo, Italy}

\author{Johannes Feist}
\affiliation{Departamento de F\'isica Te\'orica de la Materia Condensada and Condensed Matter Physics Center (IFIMAC), Universidad Aut\'onoma de Madrid, E-28049 Madrid, Spain} 

\author{Peter Rabl}
\affiliation{Vienna Center for Quantum Science and Technology,
	Atominstitut, TU Wien, 1020 Vienna, Austria}

\begin{abstract}
We address the fundamental question whether or not it is possible to achieve conditions under which the coupling of a single dipole to a strongly confined electromagnetic vacuum can result in non-perturbative corrections to the dipole's ground state. To do so we consider two simplified, but otherwise rather generic cavity QED setups, which allow us to derive analytic expressions for the total ground state energy and to distinguish explicitly between purely electrostatic and genuine vacuum-induced contributions. Importantly, this derivation takes the full electromagnetic spectrum into account while avoiding any ambiguities arising from an \textit{ad-hoc} mode truncation. Our findings show that while the effect of confinement \textit{per se} is not enough to result in substantial vacuum-induced corrections, the presence of high-impedance modes, such as plasmons or engineered $LC$ resonances, can drastically increase these effects. Therefore, we conclude that with appropriately designed experiments it is at least in principle possible to access a regime where light-matter interactions become non-perturbative. 
\end{abstract}

\maketitle

The first theoretical prediction of the Lamb shift by Bethe in 1947~\cite{Bethe1947} was an important milestone of modern physics. Not only did he show that the quantized electromagnetic vacuum leads to an observable energy shift in the Hydrogen spectrum, but also how a finite value for this shift can be obtained from a divergent perturbation theory through renormalization. In recent years, vacuum-induced modifications of molecular properties have regained considerable attention in the context of cavity QED~\cite{Ribeiro2018,FornDiaz2019,FriskKockum2019,GarciaVidal2021,Schlawin2021,Ruggenthaler2022}, where the coupling of matter to individual electromagnetic modes is strongly enhanced by a tight confinement of the field. It has been speculated that under such \emph{ultrastrong} coupling conditions \cite{FornDiaz2019,FriskKockum2019}, the electromagnetic vacuum could change the rate of chemical reactions \cite{Hutchison2012, Thomas2016, Thomas2019, Hirai2020} or modify work functions \cite{Hutchison2013}, phase transitions \cite{Wang2014} and (super-) conductivity \cite{Thomas2019arXiv,ParaviciniBagliani2019,Appugliese2022}, even without externally driving the cavity mode. However, theoretical support for such phenomena relies mainly on the analysis of phenomenological single-mode models (see, for example,~\cite{Ribeiro2018,FornDiaz2019,FriskKockum2019,GarciaVidal2021,Schlawin2021,Ruggenthaler2022} and references therein), which ignore the coupling to an infinite number of other vacuum modes. Thus, such models \textit{per se} are incapable of making reliable predictions about the magnitude or even the sign of vacuum-induced energy shifts.

In this Letter we investigate the ground state energy shift of a single dipole due to its coupling to the electromagnetic vacuum in a confined geometry. Specifically, we focus on the two basic settings of a lumped-element $LC$-resonator and a nanoplasmonic cavity, which are representative for a large variety of ultrastrong coupling experiments~\cite{FornDiaz2019,FriskKockum2019} and allow us to perform an analytic and cutoff-independent derivation of the vacuum-induced shifts of the ground state. These shifts can further be interpreted in terms of purely electrostatic effects and genuine vacuum corrections and studied as a function of the relevant system parameters. This analysis explicitly shows that when relying on confinement only, the resulting energy shifts are dominated by electrostatic corrections, while contributions from dynamical modes remain perturbative at most. However, the effect of vacuum fluctuations can be strongly enhanced for electromagnetic modes with a high impedance, where the ratio between the electric and the magnetic field strength is modified compared to free space. This condition can be reached, for example, with appropriately designed lumped-element resonators or with plasmonic resonances, where the matter component contributes a large kinetic inductance. Indeed, we find that in both of the considered settings, non-perturbative shifts of the ground state, which are comparable to the bare transition frequency of the dipole, are feasible in principle. These predictions thus serve as an important guideline for designing experiments that can reach this regime as well as a benchmark for more refined models and numerical simulations of ultrastrong coupling physics. 

\begin{figure}[b]	
	\centering
        \includegraphics[width=\linewidth]{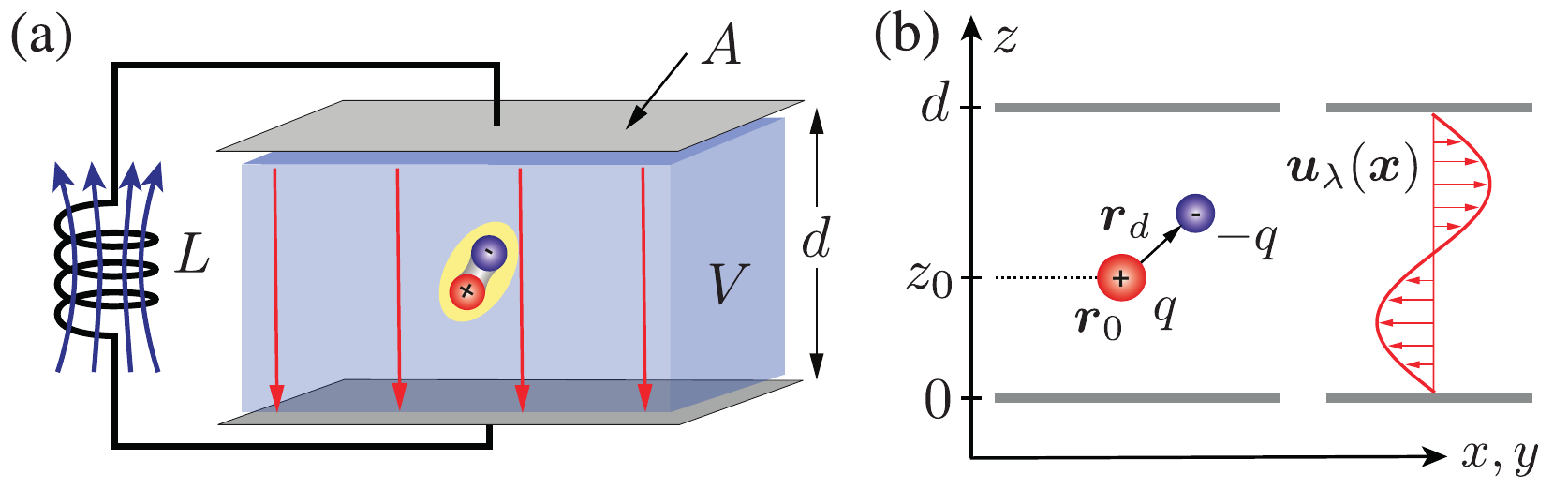}
	\caption{(a) Sketch of a generic cavity QED system with a single dipole located in a strongly confined volume $V$ between two metallic plates. The inductance $L$ is used to model an additional external cavity mode with frequency $\omega_c=1/\sqrt{LC}$ and impedance $Z=\sqrt{L/C}$, where $C=\epsilon_0 A/d$.  (b) Coordinates of the two charges representing the dipole and sketch of a mode function of the transverse vector potential. See text and \cite{Supp} for more details.}
	\label{fig:setup}
\end{figure}

\emph{QED in a confined geometry.}---We first consider the setup shown in \autoref{fig:setup}, where a single dipole is coupled to the quantized electromagnetic field in a volume $V$ defined by two perfect metallic plates with area $A$ and spacing $d$. Without loss of generality, we model the dipole as an effective particle of charge $-q$ and mass $m$, which is displaced by ${\br}_d=(x_d,y_d,z_d)$ from the opposite charge at a fixed position ${\br}_0=(0,0,z_0)$. The quantized field is represented, first of all, by an infinite set of transverse modes with frequencies $\omega_{\lambda}$ and mode functions ${\bu}_{\lambda}$, which are determined by a vanishing potential at the boundaries, $\phi({\bx})|_{\partial V}=0$. Further, we allow charges to flow freely between the two plates, which gives rise to one additional independent degree of freedom to account for a finite potential difference across the plates. This purely longitudinal mode represents our cavity mode of interest and can be modelled as an $LC$ resonance with inductance $L$, capacitance $C=\epsilon_0 A/d$ and frequency $\omega_c=1/\sqrt{LC}$. Therefore, in this setting, which is most relevant for cavity QED systems in the GHz and THz regime, the properties of the cavity mode can be engineered independently of the confinement geometry.
 
In the non-relativistic regime and under the validity of the dipole approximation, the full Hamiltonian describing this setup can be written as \cite{Supp}
\begin{equation}\label{eq:Hfull_LC}
\begin{split}
H &=  \sum_\lambda \hbar \omega_\lambda a^\dag_\lambda a_\lambda   + \hbar \omega_c a_c^\dag a_c  +H_{\rm dip}^0  +V_{\rm im}({\br}_0,{\br}_d) \\ 
&   + \frac{q}{m} \bA ({\br}_0) \cdot \bp + \frac{q^2}{2m} \bA^2({\br}_0) + \hbar g (a_c+a_c^\dag) \mu  + \frac{\hbar g^2}{\omega_c}\mu^2. 
\end{split}
\end{equation}
Here, the first two terms represent the transverse electromagnetic modes and the $LC$ cavity mode with bosonic annihilation operators $a_\lambda$ and $a_c$, respectively. The third term, $H_{\rm dip}^0= \sum_j E_j |j\rangle\langle j|$, is the Hamiltonian of the dipole in free space, which we write in terms of its eigenstates $|j\rangle$ and eigenenergies $E_j$.  

The confinement modifies the electromagnetic surrounding seen by the dipole, which, first of all, leads to a modification of the Coulomb potential by the metallic plates. In \autoref{eq:Hfull_LC}, this effect is included through $V_{\rm im}({\br}_0,{\br}_d)$, which accounts for the additional interactions between the dipole and its image charges. Secondly, the boundaries modify the frequencies and mode functions of the transverse electromagnetic modes inside the volume $V=dA$ enclosed by the plates. In the Coulomb gauge, these modes couple to the momentum ${\bp}=-\ii \hbar {\bf \nabla}_{\br_d}$ via the usual minimal coupling substitution. This gives rise to the ${\bp}\cdot \bA$ and $\bA^2$ contributions in the second line of \autoref{eq:Hfull_LC}, where
\begin{equation} \label{eq:Aopexpansion}
	\bA ({\bx})=  \sum_{\lambda} \sqrt{\frac{\hbar}{2 \epsilon_0 \omega_\lambda}}
	\left[ {\bu}_\lambda({\bx})  a_\lambda + {\bu}^*_\lambda({\bx}) a^\dag_\lambda \right] 
\end{equation}
is the vector potential. Note that, for a given geometry, the transverse mode functions, ${\bu}_\lambda$, and the electrostatic modifications, $V_{\rm im}$, are related by the  electromagnetic Green's function and cannot be treated independently of each other. Explicit expressions for ${\bu}_\lambda$ and $V_{\rm im}$ for the considered setup are given in~\cite{Supp}.


Finally, the dipole couples to the homogeneous electric field associated with the $LC$ resonance, which is represented by the last two terms in \autoref{eq:Hfull_LC}. Here, we have defined the dimensionless dipole transition operator $\mu=z_d/a_0$, where $a_0=|\langle 0| z_d|1\rangle|$ is the characteristic size of the dipole. The relevant coupling parameter can then be written as~\cite{DeBernardis2018PRA97}
\begin{equation}\label{eq:CouplingParameter}
\eta = \frac{g}{\omega_c} = \frac{q a_0}{e d} \sqrt{2 \pi \alpha \frac{Z}{Z_{\rm vac}}} ,   
\end{equation}
where $\alpha\simeq 1/137$ is the fine-structure constant, $Z=\sqrt{L/C}$ is the resonator impedance and  $Z_{\rm vac} = 1/(\epsilon_0 c)$ is the impedance of free space. This way of expressing the coupling is convenient as it immediately shows that, although being intrinsically weak, light-matter interactions can be substantially enhanced by engineering modes with a high impedance, $Z\gg Z_{\rm vac}$. Such modes necessarily correspond to longitudinal modes and involve, for example, moving charges with a large kinetic inductance.

\emph{Ground state energy shift.}---A central quantity of interest in ultrastrong-coupling cavity QED is the change in the ground state energy, $E_{\rm GS}$, when the dipole is placed inside the cavity. For $\eta \lesssim 1$ this shift can be calculated by starting from the unperturbed ground state, $|j=0\rangle|{\rm vac}\rangle$, and treating all corrections in second-order perturbation theory. 
However, due to an infinite number of modes $a_\lambda$, the result of such a calculation will diverge or contain an explicit cutoff dependence when the number of modes is restricted~\cite{Hoffmann2020,Rokaj2022}. To resolve this problem one must keep in mind that a similar divergence also occurs in free space, such that the observable energy difference $\Delta E_{\rm GS}=E_{\rm GS}|_{\rm cavity}-E_{\rm GS}|_{\rm free} $ remains finite. In~\cite{Supp} we present the details of this calculation, which results in a total energy shift of the form
\begin{equation}\label{eq:DeltaE}
\Delta E_{\rm GS}= \Delta E_{\rm im} +  \Delta E_{{\bA}}+ \Delta E_{\rm cav}.
\end{equation} 
The individual contributions account for the purely electrostatic corrections and the genuine vacuum shifts from the transverse modes and the cavity mode, respectively.  To make explicit predictions, we will evaluate these contributions for an isotropic harmonic dipole with frequency $\omega_0$. However, up to numerical prefactors, all results apply to other systems with the same characteristic transition frequency as well.   

\begin{figure}[tb]	
	\centering
	\includegraphics[width=\linewidth]{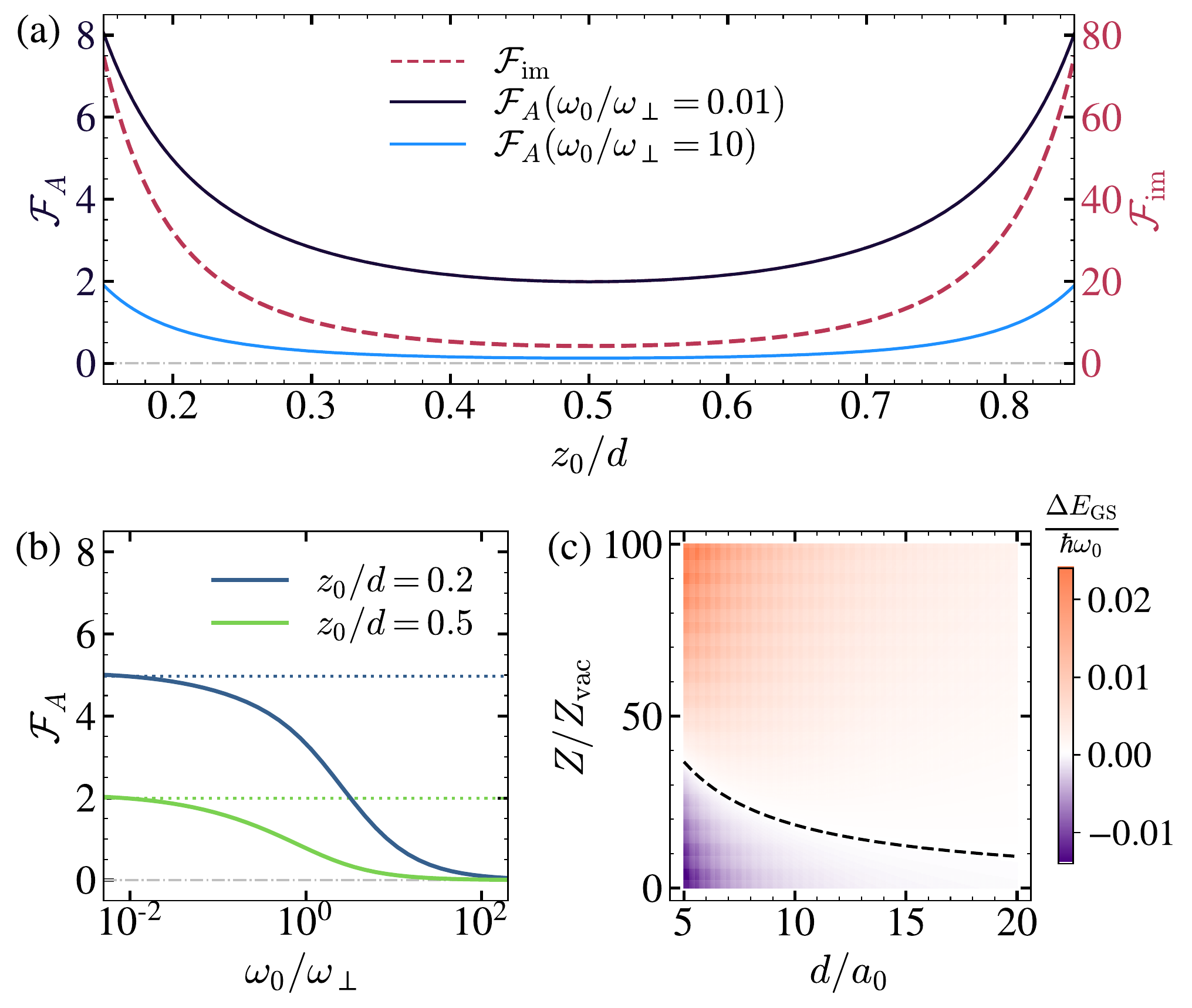}
	\caption{(a) Dependence of the dimensionless functions $\mathcal{F}_{\rm im}$ and $\mathcal{F}_{\bA}$ on the dipole position within the cavity, $z_0/d$. (b) Dependence of $\mathcal{F}_{\bA}$ on the ratio $\omega_0/\omega_\perp$, where $\omega_\perp=\pi c/d$, for two different positions $z_0/d$. (c) Total value of the ground state energy shift $\Delta E_{\rm GS}$ when the dipole is coupled to an $LC$ resonance with impedance $Z$ and the plates are separated by a distance $d$. For this plot a value of $z_0/d = 0.5$,  $\hbar\omega_0=V_C/2$, and $q=e$ have been assumed. }
	\label{fig:XFigure_es_factorFim}
\end{figure}

In the limit $\sqrt{A}\gg d$, the image charge potential $V_{\rm im}$ can be evaluated analytically and the resulting shift can be approximately written as~\cite{Supp} 
\begin{equation}
	\Delta E_{\rm im}  \simeq  - V_C \frac{a_0^3}{d^3} \mathcal{F}_{\rm im}(z_0/d),
\end{equation}
where $V_C= q^2/(4\pi \epsilon_0 a_0)$ is the relevant Coulomb energy. The dimensionless function
\begin{equation}
\mathcal{F}_{\rm im}(x)= \frac{1}{8} \left[ \frac{2}{x^3}-\Psi^{(2)}\left(1-x\right) - \Psi^{(2)}\left(1+x\right) \right],
\end{equation}
where $\Psi^{(2)}(x)$ is the polygamma function, is plotted in \autoref{fig:XFigure_es_factorFim} (a). It assumes a minimal value of $\mathcal{F}_{\rm im}(1/2)\approx 4.21$ at the center of the cavity and scales as $\mathcal{F}_{\rm im}(z_0/d\rightarrow 0)\simeq d^3/(4z_0^3)$ for small $z_0$, where it reproduces the usual van der Waals energy of an atom in front of a single metallic plate.  

The evaluation of the shift $\Delta E_{\bA}$ from the transverse modes is more subtle~\cite{Miloni1994} and only gives meaningful predictions once the corresponding free-space contribution is subtracted. Techniques to do so have originally been developed for calculating the Lamb shift~\cite{Bethe1947} and Casimir-Polder forces~\cite{CasimirPolder1948,Dzyaloshinskii1961,Wylie1984}, but have later been applied to cavity geometries as well~\cite{Casimir1948,Dzyaloshinskii1961,Barton1970,Power1982,Svozil1985,Reiche2020}.  Our analytic calculations of $\Delta E_{\bA}$ in~\cite{Supp} follow closely the derivations presented in Ref.~\cite{Barton1970}, but extended to the relevant limit of strong confinement, $\omega_0/\omega_\perp < 1$, where $\omega_\perp =\pi c/d$ is the frequency of the fundamental transverse mode and $c$ is the speed of light. In addition, we use numerical summation to verify that all results are independent of the precise details of the cutoff function and already converge for rather low values of the cutoff scale $\Lambda\approx 5 \omega_\perp$. This is a crucial observation, since it shows that none of the following results depends on the often unknown high-frequency properties of the model. From this analysis we obtain~\cite{Supp} 
\begin{equation}\label{eq:DeltaEA}
\Delta E_{\bA} = \alpha \hbar \omega_0 \left(\frac{q a_0}{ed}\right)^2     \mathcal{F}_{\bA}\left(\frac{\omega_0}{\omega_\perp},\frac{z_0}{d}\right),
\end{equation}
where the dimensionless scaling function $\mathcal{F}_{\bA}>0$ is plotted in \autoref{fig:XFigure_es_factorFim} (a). For $\omega_0\rightarrow 0$ it assumes a value of $\mathcal{F}_{\bA}(0,1/2) =2\pi/3$ at the center of the cavity and scales as  $\mathcal{F}_{\bA}(0,z_d/d\rightarrow 0) \simeq d^2/(2\pi  z_0^2)$ near the plate. For $\omega_0>\omega_\perp$, i.e., when retardation effects become important, this function decreases rapidly, as shown in \autoref{fig:XFigure_es_factorFim} (b).

From Eq.~\eqref{eq:DeltaEA} we see that compared to electrostatic effects, the overall energy shift resulting from all transverse modes is positive. This is an important finding and shows that for strong confinement, the positive energy correction from the $\bA^2$-term dominates over the negative second-order shifts obtained from the $\bp\cdot\bA$ coupling.
However, compared to electrostatics, the magnitude of $\Delta E_{\bA}$ is suppressed by the fine-structure constant and we obtain the bound
\begin{equation} \label{eq:Bound}
\begin{split}
\frac{\Delta E_{\bA}}{|\Delta E_{\rm im} |} =\,&  \alpha   \frac{\hbar \omega_0}{V_C} \frac{d}{a_0}  \frac{  \mathcal{F}_{\bA}(\omega_0/\omega_\perp,z_0/d)}{\mathcal{F}_{\rm im}(z_0/d)}\\
=\,& \pi \frac{\omega_0}{\omega_\perp} \frac{  \mathcal{F}_{\bA}(\omega_0/\omega_\perp,z_0/d)}{\mathcal{F}_{\rm im}(z_0/d)} < 1,
\end{split}
\end{equation} 
which holds in all parameter regimes~\cite{Supp}. This observation is consistent with the fact that Casimir-Polder forces between a dipole and a metallic plate are attractive and shows that even an extreme confinement, $d\sim a_0$, does not change this result. \autoref{eq:DeltaEA} also implies that for any $d>a_0$ the combined vacuum shift resulting from all transverse modes is only a small fraction of the bare transition frequency, $\omega_0$. 

%

The third term in  \autoref{eq:DeltaE} arises from the coupling of the dipole to the additional $LC$ resonance. This mode is absent in free space and therefore we can use standard perturbation theory to obtain
\begin{equation}
\Delta E_{\rm cav}= -\frac{\hbar g^2}{\omega_c+\omega_0} + \frac{\hbar g^2}{\omega_c} .
\end{equation}
In the limit $\omega_0\ll \omega_c$ and using~\autoref{eq:CouplingParameter}, we can write this contribution as
\begin{equation}\label{eq:DeltaEcavity}
\Delta E_{\rm cav}\simeq\alpha \hbar \omega_0 \left(\frac{q a_0}{ed}\right)^2 \frac{Z}{Z_{\rm vac}} \mathcal{F}_{\rm cav},
\end{equation}
with a numerical factor $\mathcal{F}_{\rm cav}=2\pi$.
Like for the transverse modes, the vacuum shift induced by the $LC$ resonance is positive and has the same scaling with frequency and distance $d$. The overall magnitude, however, is enhanced by the ratio $Z/Z_{\rm vac}$, which can be used to compensate the small value of the fine-structure constant. 

For GHz resonators, values of $Z/Z_{\rm vac}\sim 50-80$ have already been demonstrated using optimized  geometric inductors~\cite{Peruzzo2020} or superinductors~\cite{Kuzim2019}. Although experimentally challenging, \autoref{eq:DeltaEcavity} implies that under such conditions, non-perturbative vacuum shifts, $\Delta E_{\rm cav}/(\hbar \omega_0)\sim 1$, are in principle accessible. As illustrated in \autoref{fig:XFigure_es_factorFim} (c), the coupling to such high-impedance modes can also result in `anomalous' vacuum shifts, $\Delta E_{\rm GS}>0$, where the positive contribution from the dynamical modes exceeds electrostatic effects.  This is very intriguing, as such a positive shift implies an outward vacuum pressure on the cavity mirrors, in contrast to the attractive Casimir force arising from the zero-point energy of the electromagnetic modes~\cite{NoteCasimir}. However, in contrast to related previous predictions~\cite{Rokaj2022}, we find that the experimental conditions for accessing this regime are rather extreme and more refined studies will be necessary to assess the role of such effects for potential applications.  

\begin{figure}[t]	
	\centering
        \includegraphics[width=\linewidth]{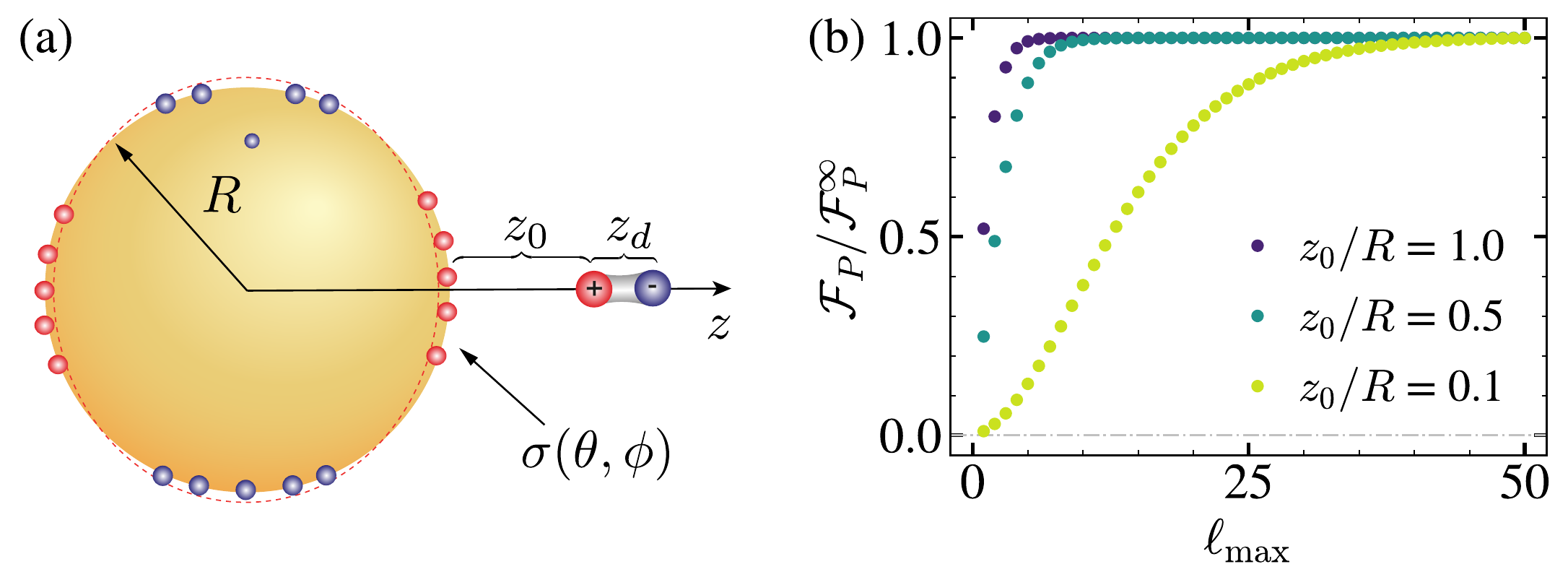}
	\caption{(a) Sketch of a dipole located a distance $z_0$ above the surface of a plasmonic nanosphere cavity with radius $R$ and surface charge density $\sigma$. (b) Plot of the function $\mathcal{F}_P$ defined in \autoref{eq:FP} versus the cutoff $\ell_{\rm max}$, where $\mathcal{F}_P^\infty$ denotes the asymptotic value for $\ell_{\rm max}\rightarrow \infty$. For these plots $\omega_0/\omega_P=1$ and different values of $z_0$ have been assumed. }
	\label{fig:plasmon}
\end{figure}

\emph{Plasmonic nanosphere cavity.}---In the previous setup we have assumed perfect metallic boundary conditions and thereby ignored dynamical electromagnetic modes associated with the redistribution of electrons inside the metal. In our second example we specifically address the influence of these excitations and consider the setup shown in \autoref{fig:plasmon} (a), where the dipole is placed at a distance $z_0$ above the surface of a plasmonic nanosphere cavity with radius $R$. Following Ref.~\cite{Prodan2004}, we model the electrons inside this sphere as an incompressible fluid of density $n_0$, which exhibits a discrete set of plasmon modes with frequencies $\omega_\ell=\omega_P \sqrt{\ell/(2\ell+1)}$. Here $\omega_P=\sqrt{e^2 n_0/(m_e\epsilon_0)}$ is the plasma frequency and $m_e$ is the electron mass.  These modes describe excitations of the surface charge density $\sigma$ with an angular momentum quantum number $\ell$. By restricting the motion of the dipole along the $z$-axis for simplicity, the Hamiltonian for this setup is~\cite{Supp}
%
\begin{equation} \label{eq:Hfull_PM}
\begin{split} 
H= &H_{\rm dip}^0  +V_{\rm im}+   \hbar \sum_{\ell=1}^{\ell_{\rm max}}  \left[ \omega_\ell a_\ell^\dag a_\ell + g_\ell (a_\ell+ a_\ell^\dag) \mu+   \frac{ g_\ell^2}{\omega_\ell} \mu^2 \right],
\end{split} 
\end{equation}
where $V_{\rm im} \equiv V_{\rm im}(z_0,z_d)$ denotes the image potential experienced by a static dipole in front of the sphere. In \autoref{eq:Hfull_PM}, $a_\ell$ ($a_\ell^\dag)$ are the annihilation (creation) operators for the plasmon modes up to a maximal quantum number $\ell_{\rm max}$ and we have defined the coupling constants $g_\ell = g_P  \sqrt[4]{\ell (\ell+1)^4/(2\ell+1)} (R/(R+z_0))^{\ell+2}/2$. Here,     
\begin{equation}
\eta_P= \frac{g_P}{\omega_P} = \frac{q a_0}{eR} \sqrt{2\pi \alpha \frac{Z_P}{Z_{\rm vac}}}
\end{equation} 
and $ Z_P=(\pi\epsilon_0 R \omega_P)^{-1}$ is the characteristic impedance. Note that \autoref{eq:Hfull_PM} has been derived starting from Coulomb interactions only, and therefore neglects small corrections from the coupling to transverse modes, which we have already evaluated above. In the limit $\ell_{\rm max}\rightarrow \infty$, the so-called $P^2$-term, $\sum_\ell g_\ell^2\mu^2/\omega_\ell $, cancels the instantaneous image potential, $V_{\rm im}$, exactly. However, by writing \autoref{eq:Hfull_PM} in this canonical form one avoids a double-counting of electrostatic interactions also for a finite number of modes and one recovers the correct single-mode cavity QED Hamiltonian for $\ell_{\rm max}=1$~\cite{DeBernardis2018PRA97}.

Since the plasmon modes are absent in free space, the ground state shift resulting from Hamiltonian \eqref{eq:Hfull_PM} is given by $\Delta E_{\rm GS}\simeq \Delta E_{\rm im} + \Delta E_{P}$. Here $\Delta E_{\rm im}$ accounts again for the electrostatic contribution $\sim V_{\rm im}$, which in the limit $z_0/R\ll1$ reduces to the van der Waals interaction~\cite{Supp}
\begin{equation}
\Delta E_{\rm im}(z_0\ll R) \simeq - \frac{q^2a_0^2}{4\pi \epsilon_0 (2z_0)^3}
\end{equation}
for a dipole oriented along the $z$-direction. The energy shift resulting from the coupling to the dynamical plasmon modes can be written as   
\begin{equation}\label{eq:DeltaEPlasmon}
\Delta E_{P}\simeq  \alpha  \hbar \omega_0  \left(\frac{q a_0}{ez_0}\right)^2   \frac{Z_{\rm eff}(z_0)}{Z_{\rm vac}}   \mathcal{F}_P\left(\frac{\omega_0}{\omega_P}, \frac{z_0}{R}\right),
\end{equation} 
where $Z_{\rm eff}(z_0)=(\pi\epsilon_0 z_0 \omega_P)^{-1}$ and
\begin{equation}\label{eq:FP}
	 \mathcal{F}_P(x,y)= \frac{\pi y^3}{2}  \sum_{\ell=1}^{\ell_{\rm max}} \frac{(\ell+1)^2  \sqrt{(2\ell+1)/\ell} }{1+x\sqrt{(2\ell+1)/\ell}} \left(\frac{1}{1+y} \right)^{2\ell+4}.
\end{equation}
In \autoref{fig:plasmon} (b) we plot the value of this function for increasing $\ell_{\rm max}$ and show that it converges to a finite value $\mathcal{F}^\infty_{P}(x,y)$ for a sufficiently large $\ell_{\rm max}$. This demonstrates that also in this scenario we can obtain an unambiguous result for the ground state energy shift that does not depend on an \emph{ad-hoc} mode truncation. However, \autoref{fig:plasmon} (b) also shows that for $z_0< R$, a single-mode cavity QED model would be insufficient to predict this shift accurately.

In \autoref{eq:DeltaEPlasmon}, we have combined the contribution from all plasmon modes into an effective impedance $Z_{\rm eff}(z_0)$, which shows that for small dipole-sphere separations, $z_0$ replaces the radius $R$ as the relevant length scale. In this limit, $\mathcal{F}^\infty_{P}(x\ll1,y\ll1 )\simeq \pi/\sqrt{32} $. For plasma frequencies in the range of $1-10$ eV~\cite{BookKittel,Naik2013} and $z_0=0.5$ nm~\cite{Chikkaraddy2016,Kuisma2022} we obtain $Z_{\rm eff}/Z_{\rm vac}\approx 12-125$, which for a large molecule with $qa_0/(ez_0)\approx 1$ corresponds to $\Delta E_{P}/(\hbar \omega_0)\approx 0.05-0.50$, still assuming that $\omega_0 \lesssim\omega_P$.  Therefore, also in the plasmonic case, non-perturbative vacuum effects are within experimental reach.  Note however, since $Z_{\rm eff}(z_0)$ is fixed by geometry and cannot be engineered independently, we find that the total shift in this setup, $\Delta E_{\rm im} + \Delta E_{P} < 0$, is negative for all values of $z_0$ and $\omega_0$~\cite{Supp}.

\emph{Summary and conclusions.}---In summary, we have presented an ab-initio and cutoff-independent derivation of the total ground state energy shift in two prototypical cavity QED settings. By focusing on simple geometries, we have obtained analytic predictions for the dependence of this shift on the most relevant experimental parameters and provided a clear distinction between purely electrostatic corrections and genuine vacuum effects. Our results demonstrate that the coupling to strongly confined transverse modes, as often considered in cavity QED, is not enough to induce significant perturbations of the ground state. The main effect of the confinement are electrostatic modifications, which in turn are often neglected in this context. These findings are consistent with the previous literature on Casimir-Polder interactions and establish a direct link between those closely related, but so far largely disconnected fields of research.

However, going beyond such conventional settings, our analysis predicts a significant enhancement of vacuum corrections in the presence of high-impedance modes. Under such conditions, the regime of non-perturbative cavity QED~\cite{DeBernardis2018PRA97}, with its many intriguing phenomena~\cite{Ridolfo2013,Galego2015,Flick2017,MartinezMartinez2018,Schlawin2019,Rivera2019,Schuler2020,Pilar2020,Li2020,Ashida2021,Couto2022}, comes within experimental reach. The current very general analysis can help to guide further experimental and theoretical progress in this direction.

\emph{Acknowledgments.}---This work was supported by the Austrian Science Fund (FWF) through Grant No. P31701 (ULMAC), Grant No. P32299 (PHONED) and the ESPRIT fellowship ESP 230-N (mesoQED).

\bibliography{bibLamb} 
\bibliographystyle{apsrev4-2} 

\cleardoublepage

\onecolumngrid
\begin{center}
	\textbf{SUPPLEMENTARY MATERIAL}
\end{center}

\twocolumngrid

\section{Cavity QED Hamiltonian for the $LC$-resonator setup}
\label{sec:SMcavityQEDHamiltonianLC}

In this section we discuss in more detail the derivation of the cavity QED Hamiltonian for the setup shown in \autoref{fig:setup} in the main text. Here, a single dipole is located between two perfectly conducting metallic plates, which form the capacitance of a lumped-element $LC$ circuit. The capacitor plates are located at positions $z=0$ and $z=d$, and enclose the quantization volume $V=Ad$, where $A=l_x l_y$ is the area of the plates. Throughout this section we will assume that $l_x=l_y=l$ and $d\ll l$, such that boundary effects in the $x-y$ plane can be neglected.

\subsection{Dynamical degrees of freedom and classical equations of motion} 

While the following analysis can be readily generalized to more complex matter systems, we model the molecular dipole as a single particle of charge $-q$ and mass $m$, which is displaced by ${\br}_d=(x_d,y_d,z_d)$ from an opposite charge $+q$ at a fixed location ${\br}_0=(0,0,z_0)$.  The corresponding charge and current densities are given by 
\begin{eqnarray} \label{SM1_densityDist}
\rho({\bx},t) &=&    q  \left[ \delta ({\bx}- {\br}_0)- \delta({\bx}-  {\br}_0 -{\br}_d(t))\right],\\
 {\bj}({\bx},t)&=& -  q  \dot {\br}_d(t) \delta ({\bx}- {\br}_0- {\br}_d(t)),
\end{eqnarray}
and obey the continuity equation, $ \partial_t \rho + {\bf \nabla} \cdot {\bj}=0$. The classical equation of motion for the charged particle is determined by the Lorentz force,
\begin{equation}
m\ddot{\br}_d =-q \left[ {\bE}({\br}_0+{\br}_d) + \dot {\br}_d  \times {\bB}({\br}_0+{\br}_d) \right],
\end{equation}
where ${\bE}$ and ${\bB}$ are the total electric field and the total magnetic field, respectively. Within the volume $V$, the electric and magnetic fields can be written as
\begin{gather}
	{\bE} = - \frac{\partial {\bA}}{\partial t} - \nabla \phi-  \frac{U}{d} \boldsymbol{e}_z, \qquad  {\bB} = \nabla \times {\bA} .
\end{gather}
Here, $\phi$ and ${\bA}$ represent the scalar and vector potential between the metallic plates and for the boundary conditions $\phi|_{z=0}=\phi|_{z=d}=0$. In addition, we allow charges to flow on and off the metallic plates, which creates an additional potential difference $U$ along the $z$-direction (where $\boldsymbol{e}_z$ is the unit vector along this direction). We neglect the effect of the magnetic fields produced by this external currents on the dynamics of the dipole. 

\subsubsection{Confined electromagnetic fields}  
In the Coulomb gauge, $\nabla \cdot {\bA} = 0$, the scalar and vector potentials obey   
\begin{gather} \label{eq:SM1_PoissonEq}
	\begin{aligned}
		&\nabla^2 \phi = -\frac{\rho}{\epsilon_0},\qquad &\left( \frac{1}{c^2}\frac{\partial^2 }{\partial t^2} - \nabla^2 \right) {\bA } =  \frac{{\bj^\perp}}{\epsilon_0 c^2},
	\end{aligned}
\end{gather}
where $c$ is the speed of light, $\epsilon_0$ is the vacuum permittivity, and
\begin{equation}
{\bj}^\perp= {\bj} -  \epsilon_0 \nabla \frac{\partial}{\partial t} \phi
\end{equation}
is the transverse current density. 

As usual, the potential $\phi$ is not an independent dynamical degree of freedom and represents the instantaneous potential,
\begin{equation} \label{eq:SM1_phiasafunctionofG}
 \phi({\bx},t)=\frac{1}{\epsilon_0} \int \dd^3 x' \,  G({\bx},{\bx}') \rho({\bx}',t),
\end{equation}
associated with the charge distribution $ \rho({\bx}',t)$. Here, $G({\bx},{\bx}')$ is the electrostatic Green's function, which satisfies $\nabla^2 G({\bx},{\bx}')=-\delta({\bx}-{\bx}')$ and $G(z=0, {\bx}')= G(z=d, {\bx}') = 0$. In the limit $d\ll l $, this Green's function can be written as
\begin{equation}
G({\bx},{\bx}')\simeq  \frac{1}{4\pi | {\bx} - {\bx}'|} + G_{\rm im}({\bx},{\bx}'),
\end{equation} 
where $G_{\rm im}({\bx},{\bx}')$ accounts for the contribution from all the image charges in an infinite parallel plate configuration. Similarly, we can write 
\begin{equation} 
 \phi({\bx})\simeq \phi_{\rm free} ({\bx}) + \phi_{\rm im} ({\bx}),
\end{equation} 
where $\phi_{\rm free} ({\bx})$ is the unperturbed Coulomb potential in free space and $\phi_{\rm im} ({\bx})$ is the additional potential arising from the image charges. Explicit expressions for $\phi_{\rm im} ({\bx})$ will be derived below (see \autoref{sec:SMesPotentialLC}). 

For the vector potential we use the mode expansion 
\begin{equation} \label{eq:SM3_Aexpansion} 
	{\bA}({\bx},t)  = \frac{1}{\sqrt{2 \epsilon_0}}  \sum_{\lambda} 
	\left[ {\bu}_\lambda ({\bx}) \alpha_\lambda(t)  + {\bu}_\lambda^*({\bx}) \alpha^*_\lambda(t) \right],
\end{equation}
where the amplitudes $\alpha_\lambda$ are the dynamical degrees of freedom and the mode functions ${\bu}_\lambda$ satisfy the Helmholtz equation 
\begin{equation}
\left( \frac{\omega_\lambda^2}{c^2}+ \nabla^2 \right){\bu}_\lambda = 0.
\end{equation}  
They are normalized to 
\begin{equation} 
\int_V \dd^3 x  \  {\bu}_\lambda ({\bx})  \cdot {\bu}_{\lambda'}^*({\bx}) = \delta_{\lambda \lambda'}.
\end{equation}
The index $\lambda\equiv ({\bk}, \sigma) $ runs over two possible polarizations ($\sigma= 1,2$) and all wavevectors ${\bk}=(k_x,k_y,k_z)$ compatible with the boundary conditions (namely, perfect metallic boundary conditions in the $z$-direction and, for simplicity, periodic boundary conditions along $x$ and $y$). Therefore,
\begin{equation} \label{eq:SM1_modesBounCond} 
	k_x = \frac{2 \pi}{l_x} n_x \ , \hspace{3mm} 
	k_y = \frac{2 \pi}{l_y} n_y \ , \hspace{3mm} 
	k_z = \frac{\pi}{d} n \ ,
	\vspace{1mm}
\end{equation}
where $n_x, n_y \in \mathbb{Z}$ and $n = 0, 1, 2 \dots$ The frequency of the mode is given by $\omega_\lambda = c k$, where $k = |\bk|$. Explicitly, the mode functions are given by
\begin{equation} \label{eq:SM3_modesModeFunctionsD3}
	{\bu}_\lambda ({\bx})  = \sqrt{\frac{2}{V (1 + \delta_{n0})}} \ \ee^{\iii (k_x x + k_y y)}
	\begin{pmatrix}
		\ii \boldsymbol{\varepsilon}_{\lambda x} \sin (k_z z)	\\
		\ii \boldsymbol{\varepsilon}_{\lambda y} \sin (k_z z) \\
		\boldsymbol{\varepsilon}_{\lambda z} \cos (k_z z)
	\end{pmatrix}  ,
\end{equation}
where $\delta_{n0}$ is the Kronecker delta accounting for the extra factor when $n=0$. The  (unit) polarization vectors 
$\boldsymbol{\varepsilon}_{\lambda} (= \boldsymbol{\varepsilon}_{{\bk} \sigma}) =  (\boldsymbol{\varepsilon}_{\lambda x}, \boldsymbol{\varepsilon}_{\lambda y}, \boldsymbol{\varepsilon}_{\lambda z})$ 
obey
\begin{equation}
	{\bk} \cdot \boldsymbol{\varepsilon}_{{\bk},\sigma} = 0, \hspace{5mm}
	\boldsymbol{\varepsilon}_{{\bk},\sigma}^* \cdot \boldsymbol{\varepsilon}_{{\bk},\sigma'} = \delta_{\sigma \sigma'}. 
\end{equation}

\subsubsection{$LC$ resonance} 
By allowing charges to flow on and off the metallic plates, we obtain the potential difference $U$ as an additional dynamical degree of freedom, which is independent of the vector and scalar potentials introduced above. In the GHz and THz regime this degree of freedom can be generically modelled as a lumped-element $LC$ resonator, with a capacitance $C=\epsilon_0 A/d$ formed by the two metallic plates and a system specific inductance $L$. By denoting by $Q$ the total charge on the upper capacitor plate and by $\Phi$ the magnetic flux in the inductor, the relevant circuit equations of motion are
\begin{equation}	\label{eq:SM1_LCcircuit_timePhi}
	\dot \Phi  = U= \frac{Q-Q_{\rm ind}}{C},\qquad 
	\dot Q  = - \frac{\Phi}{L}.
\end{equation}
Here, $Q_{\rm ind}$ is the charge on the upper plate that is induced by the dipole and thus does not contribute to the potential difference $U$. For a large aspect ratio $l\gg d$ and sufficiently far away from the edges of the metallic plates, this induced charge is well approximated by~\cite{Takae2013_SM}
\begin{equation}
Q_{\rm ind}= \frac{z_d q }{d} .
\end{equation} 
The resulting equation of motion for the magnetic flux is thus given by~\cite{DeBernardis2018_SM}
\begin{equation}
C \ddot \Phi+ \frac{1}{L} \Phi \simeq - \frac{q \dot z_d}{d}.
\end{equation}

\subsection{Quantization}

Starting from the equations of motion for the dipole and all electromagnetic degrees of freedom we obtain the Hamiltonian for this setup following the usual canonical quantization approach. First, the Lagrangian of the system reads
\begin{equation}
\begin{split}
L = \frac{1}{2} C \dot \Phi^2 - \frac{1}{2L} \Phi^2 + \dot\Phi Q_{\rm ind} + \frac{1}{2} m \dot \br_d^2 + \int \dd^3 x  \, \mathcal{L}_{\rm em} (\bx)
\end{split}
\end{equation}  
with the Lagrange density
\begin{equation}
\begin{split}
\mathcal{L}_{\rm em} (\bx) =&\frac{\epsilon_0}{2} \dot{\boldsymbol{A}}^2(\bx, t) - \frac{1}{2\mu_0} \boldsymbol{B}^2(\bx, t)\\
&+   {\bj^\perp} (\bx, t) \cdot \bA (\bx, t)   - \frac{1}{2}\rho (\bx, t) \phi (\bx, t).
\end{split}
\end{equation}
This yields the following canonical momenta for the system variables
\begin{align}
	\pi_\Phi  & = \frac{\partial L }{\partial \dot \Phi} = C \dot \Phi + Q_{\rm ind} =Q, \\
	\pi_{\br_d} & = \frac{\partial L}{\partial \dot {\br}_d} = m\dot \br_d - q \bA (\br_0 + \br_d) = \bp, \\
        \pi_\bA  &= \frac{\partial \mathcal{L}_{\rm em} }{\partial \dot \bA} =  \epsilon_0 \dot \bA =-\epsilon_0 \bE_\perp.
\end{align}
By promoting these classical variables and their conjugate momenta to operators satisfying $[\Phi, Q] = \ii \hbar$ and $[\br_d, \bp] = \ii \hbar$, we obtain the Hamiltonian operator for this system,
\begin{equation}
	\begin{split}
      H & =  \frac{(\bp+q\bA)^2}{2m} + V + \frac{(Q-Q_{\rm ind})^2}{2C} + \frac{\Phi^2}{2L}  \\
	 & + \int \dd^3 x \ \left\{\frac{\epsilon_0}{2} \boldsymbol{E}_\perp^2(\bx, t) + \frac{1}{2\mu_0} \boldsymbol{B}^2(\bx, t) \right\},
	\end{split}
\end{equation}
where 
\begin{equation}
V  = \frac{1}{2} \int \dd^3 x \, \rho (\bx) \phi (\bx)= V_{\rm free}+V_{\rm im}.
\end{equation}
is the total Coulomb energy. 

We proceed by expressing the quantized radiation field in terms of the usual annihilation and creation operators $a_\lambda$ and $a^\dag_\lambda$. For the $LC$ resonator we introduce the corresponding operators $a_c$ and $a_c^\dag$ via
\begin{equation}
Q  = -\sqrt{\frac{\hbar \omega_{\rm c} C}{2}} (a_c^\dagger + a_c) \ ,
	\qquad
\Phi =  \ii \sqrt{\frac{\hbar}{2 C \omega_{\rm c}}} (a_c^\dagger -a_c)  \ ,
\end{equation}
finally yielding
\begin{equation}
\begin{split}
H & = \sum_\lambda  \hbar \omega_\lambda a_\lambda^\dagger a_\lambda + \hbar \omega_c a_c^\dagger a_c  + H_{\rm dip}^0+ V_{\rm im} 
\\
& 
+\frac{q}{m} \bA (\br_0) \cdot \bp + \frac{q^2}{2m}\bA^2 (\br_0)\\
&+ \frac{q}{d} \sqrt{\frac{\hbar \omega_c}{2C}} z_d  (a_c+ a_c^\dagger) + \frac{q^2}{2d^2C} z_d^2,
\end{split}
\end{equation}
where we have already made the dipole approximation $\bA (\br_0+\br_d)\approx \bA (\br_0)$ and defined the bare dipole Hamiltonian as $H_{\rm dip}^0 = \frac{\bp^2}{2m} +V_{\rm free}$.

In a final step we introduce the dimensionless dipole transition operator $\mu =  z_d/ a_0$, where $a_0 = |\langle 0 | z_d | 1 \rangle |$,  and define the coupling constant
\begin{equation} \label{eq:SMcQEDHam_gdef}
\hbar g = \frac{q a_0}{d} \sqrt{\frac{\hbar \omega_c}{2C}} \ .
\end{equation}
By considering the impedance of the circuit, $Z=\sqrt{L/C}$, the relevant  dimensionless coupling parameter is given by~\cite{DeBernardis2018PRA97}
\begin{equation} \label{eq:SMcQEDHam_etaexp}
\eta = 	\frac{g}{\omega_{\rm c}} = \frac{q a_0}{e d} \sqrt{2 \pi \alpha \frac{Z}{Z_{\rm vac}} } \ , 
\end{equation}
where $\alpha = e^2 Z_{\rm vac} /(4 \pi \hbar)$ is the fine-structure constant and $Z_{\rm vac} = 1/(\epsilon_0 c)$  is the impedance of free space.

\section{Electrostatic potential for the parallel plate configuration}
\label{sec:SMesPotentialLC}

For the setup examined in \autoref{sec:SMcavityQEDHamiltonianLC}, here we compute the modification of the potential seen by the dipole due to the charges induced on the metallic plates. 
For the particular case of a point charge $q$ at position $\br'$, 
$\rho(\br) = q \delta(\br - \br')$ the electrostatic potential between the plates (\autoref{eq:SM1_phiasafunctionofG}) reduces to
\begin{equation} \label{eq:SM1_esrhoasafunctionofG}
	\phi^{(q)}(\br; \br')=\frac{q}{\epsilon_0}  G({\br},{\br'})  \ ,
\end{equation}
where $G(\br, \br')$ is the Green's function satisfying 
\begin{equation} \label{eq:SM1_eqforGreensfunction}
\nabla^2 G({\br},{\br'})=-\delta({\br}-{\br'})
\end{equation}
and with the boundary condition $G(z=0, \br') = G(z=d, \br') = 0$ at the plates. In the other two directions we assume periodic boundary conditions, $G(x+l_x, y+l_y, z) = G(x,y,z)$, which can be enforced by changing to Fourier space, that is,
\begin{equation} \label{eq:SM1_Green}
G({\br},{\br'}) = \frac{1}{l_xl_y}\sum_{k_x, k_y}  \  \mathcal{G}_k ({z},{z'})  \e^{-\iii k_x (x-x')} \e^{-\iii k_y (y-y')}  ,
\end{equation}
with $\mathcal{G}_k(z=0, z') = \mathcal{G}_k (z=d, z') = 0$ and $k_x$ and $k_y$ as given in \autoref{eq:SM1_modesBounCond}. The substitution of this expression into \autoref{eq:SM1_eqforGreensfunction} yields
\begin{equation*}
	\left( k_\parallel^2 - \frac{\partial^2}{\partial z^2}\right) \mathcal{G}_{k} (z, z')
	= \delta(z - z') \ ,
\end{equation*}
where $k_\parallel^2 \equiv k_x^2 + k_y^2$. The solution to this equation can be written as 
\begin{equation} \label{eq:SM1_Greenk_exp1}
	\begin{split}
	\mathcal{G}_{k}& (z, z')  =  \frac{1}{2k_\parallel} \ee^{-k_\parallel |z - z'|}
	- \frac{1}{2 k_\parallel \sinh(k_\parallel d)} \\ & \left[  
	\sinh(k_\parallel z) \ \ee^{-k_\parallel (d - z')} + \sinh[k_\parallel(d-z)] \ \ee^{-k_\parallel z'}
	\right] ,
	\end{split}
\end{equation}
that is, the sum of the homogeneous solution (which represents the regular Coulomb interaction) and an additional term that follows from the boundary conditions (and is thus associated with the charges induced on the metallic plates). By considering the expansion ${1}/{\sinh(x)} = 2 \sum_{n=0}^\infty   \e^{-x (2n+1)}$, the Green's function reduces to \cite{Takae2013_SM}
\begin{equation} \label{eq:SM1_Greenk_expgoal}
		\mathcal{G}_{k} (z, z')  =  \frac{1}{2 k_\parallel }  \sum_{n=-\infty}^\infty \left( 
		\e^{-k_\parallel |z  - z' -2dn|}  -	\e^{ -k_\parallel|z +z' - 2dn|} 
		\right)  .
\end{equation}
Substituting this expression into \autoref{eq:SM1_Green} we obtain
\begin{equation} 
	\begin{split}
	G({\br},{\br'}) = \frac{1}{l_xl_y}  \sum_{k_x, k_y} \sum_{n=-\infty}^\infty  \ 
\frac{1}{2 k_\parallel }  \left( 
\e^{-k_\parallel |z  - z' -2dn|} \right. \\ \left. -	\e^{ -k_\parallel|z +z' - 2dn|} 
\right) 	
	\e^{-\iii k_x (x-x')} \e^{-\iii k_y (y-y')}  ,
		\end{split}
\end{equation}
Finally, for $l_x=l_y=l\gg d$, we replace the sums over the $k$-vectors by integrals and end up with 
\begin{equation} 
	\begin{split}
		G({\br},{\br'}) & =  \frac{1}{4 \pi}  \sum_{n=-\infty}^\infty
		\left( \frac{1}{ |\br  - \br' + 2nd\boldsymbol{e}_z|} \right.
		\\  &  \left.
		- \frac{1}{|\br  - \br' + (2nd + 2z')\boldsymbol{e}_z|} \right).
	\end{split}
\end{equation}
If we substitute this expression into \autoref{eq:SM1_esrhoasafunctionofG}, we obtain the electrostatic potential associated with this configuration,
\begin{equation}  \label{eq:SM1_esPotentialforq}
	\begin{split}
		\phi^{(q)}(\br; \br') =  \frac{q}{4 \pi \epsilon_0}  \sum_{n=-\infty}^\infty  &
		\left( \frac{1}{ |\br  - \br' + 2nd\boldsymbol{e}_z|} \right.
		\\  &  \left.
		- \frac{1}{|\br  - \br' + (2nd + 2z')\boldsymbol{e}_z|} \right)  .
	\end{split}
\end{equation}
This corresponds to the potential created by an infinite collection of charges $+q$ located at $z=z'-2nd$ and an infinite collection of charges $-q$ at $z=-z'-2nd$, with $n$ running over all integers. Notice that the electrostatic potential in free space would reduce to the Coulomb expression
\begin{equation}  \label{eq:SM1_esfreePotentialforq}
	\phi_{\rm free}^{(q)} (\br; \br') = \frac{q}{4 \pi \epsilon_0}  
	\frac{1}{|\br - \br'|} \ .
\end{equation}

\begin{figure}[t] 
	\centering
	\includegraphics[width=\linewidth]{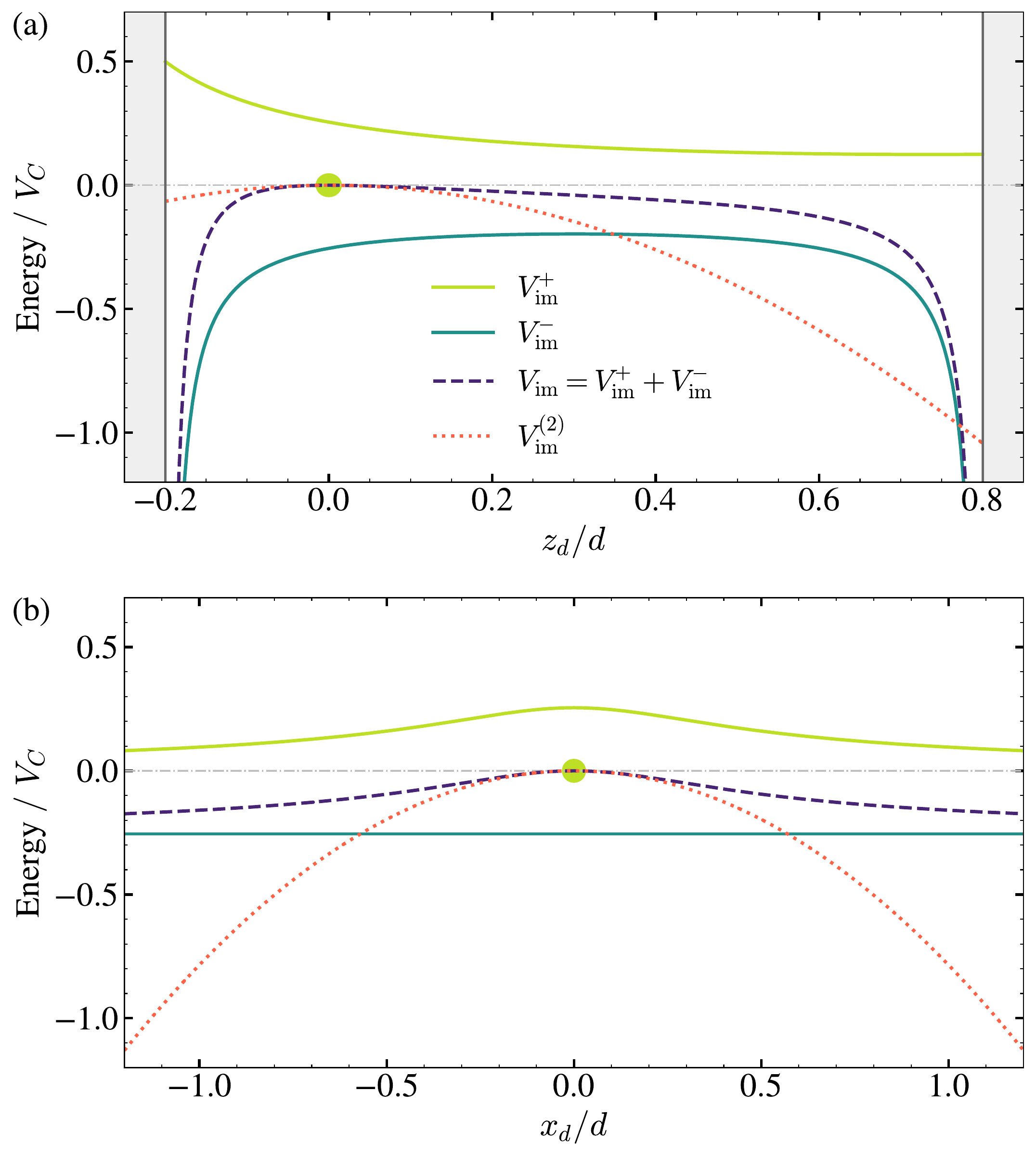}
	\caption{Modification of the electrostatic energy due to the image charges associated with a molecular dipole placed between two infinite parallel plates. Energy is plotted in units of $V_C = q^2 /( 4 \pi \epsilon_0 a_0)$, where $a_0 \equiv |\langle 0 |z_d | 1 \rangle|$ is the characteristic size of the dipole. The plates are separated by a distance $d=10a_0$ and the positive charge is located at $z_0/d = 0.2$. The energy is plotted as a function of $z_d$ when $x_d= y_d = 0$ (a) and as a function of $x_d$ when $y_d=z_d=0$ (b). The parts ascribed to the positive and negative charges, $V_{\rm im}^+$ and $V_{\rm im}^-$, are plotted separately with yellow and turquoise continuous lines, respectively. The total contribution $V_{\rm im} = V_{\rm im}^+ + V_{\rm im}^-$ is represented with a dashed purple line, where the exact numerical result $V_{\rm im}$ is compared to the analytical expression $V^{(2)}_{\rm im}$ obtained from perturbation theory (dotted red line).}
	\label{fig:FigureSM_esMod_Vim}
\end{figure}

In the setup considered here, a positive charge $+q$ is fixed at position $\br_0 = (0, 0, z_0)$ while the opposite charge $-q$ is displaced by $\br_d = (x_d, y_d, z_d)$, resulting in the total potential 
\begin{equation}  \label{eq:SM1_esPotentialforSetup}
	\phi (\br) = \phi^{(q)} (\br; \br_0) + \phi^{(-q)} (\br; \br_0 + \br_d),
\end{equation}
where each of these terms are of the form given in \autoref{eq:SM1_esPotentialforq}.
We are interested in the modification of the electrostatic energy due to the presence of the metallic plates, which  is nothing but its difference with respect to free space, 
\begin{equation}  \label{eq:SM1_esEnergyVim}
	V_{\rm im} = \frac{1}{2} \int \dd^3 r  \rho (\br) \left[ \phi (\br) - \phi_{\rm free} (\br)  \right] .
\end{equation}
Therefore, taking into account the charge density in \autoref{SM1_densityDist} and the electrostatic potential in \autoref{eq:SM1_esPotentialforSetup}, the electrostatic energy associated with the image charges becomes
\begin{equation} \label{eq:SM1_Vimexp0}
	\begin{split}
	V_{\rm im}   = \frac{q}{2} & \left[
	 \phi^{(q)} (\br_0, \br_0) 	- 	 \phi^{(q)} (\br_0 + \br_d, \br_0) 
\right. \\  & \left. +  \phi^{(-q)} (\br_0, \br_0 + \br_d)
 -  \phi^{(-q)}  (\br_0 + \br_d, \br_0 + \br_d)
	\right]
	\\
	- \frac{q}{2}& \left[
	\phi^{(q)}_{\rm free} (\br_0, \br_0) 	- 	 \phi^{(q)}_{\rm free} (\br_0 + \br_d, \br_0) 
	\right. \\  & \left. +  \phi^{(-q)}_{\rm free} (\br_0, \br_0 + \br_d)
	-  \phi^{(-q)}_{\rm free}  (\br_0 + \br_d, \br_0 + \br_d)
	\right] .
		\end{split}
\end{equation}
The explicit expression of these terms can be taken from \autoref{eq:SM1_esPotentialforq} and \autoref{eq:SM1_esfreePotentialforq}, yielding
\begin{equation} \label{eq:SM1_Vimexp1}
	\begin{split}
	V_{\rm im}  =  \frac{q^2}{4 \pi \epsilon_0}  & \left[
	\sum_{n\neq 0} \frac{1}{|2nd|}
	- \frac{1}{2} \sum_{n}
	\frac{1}{|2z_0+2nd|} \right. \\ 
	 & -  \frac{1}{2} \sum_{n} \frac{1}{|2z_0+2z_d+2nd|} 
	 \\
	 & - \sum_{n\neq 0} \frac{1}{\sqrt{x_d^2 + y_d^2 + (z_d+2nd)^2}} 
	 \\ 
	 &  \left. +	\sum_{n} \frac{1}{\sqrt{x_d^2 + y_d^2 + (2z_0 + z_d+2nd)^2}}
	\right].
\end{split}
\end{equation}
This expression can be decomposed into the parts associated with the positive and negative charges, $V_{\rm im} = V_{\rm im}^+ + V_{\rm im}^-$, where
	\begin{equation}
		\begin{split}
			V_{\rm im}^+  
			 = - \frac{q^2}{4 \pi \epsilon_0}   & \left[  \sum_{n=1}^\infty \frac{1}{\sqrt{x_d^2 + y_d^2 + (z_d+2nd)^2}} \right. \\
			 & + \sum_{n=1}^\infty \frac{1}{\sqrt{x_d^2 + y_d^2 + (z_d-2nd)^2}}  
			\\ 
			&  \left. - \sum_{n=-\infty}^\infty \frac{1}{\sqrt{x_d^2 + y_d^2 + (2z_0 + z_d+2nd)^2}}
			\right] 
		\end{split}
	\end{equation}
and
%
\begin{equation}
	\begin{split}
		V_{\rm im}^- 	
		=  - \frac{q^2}{4 \pi \epsilon_0 d}  & \left[
		 \frac{d}{4z_0} 
		- \sum_{n=1}^\infty \frac{1}{n}
		+ \frac{1}{2} \sum_{n=1}^\infty  \frac{n}{n^2 -(z_0/d)^2}   
		\right. \\ 	& \left.
		+  \frac{1}{4} \frac{d}{z_0+z_d} 
		+  \frac{1}{2} \sum_{n=1}^\infty \frac{n}{n^2 - [(z_0 +z_d)/d]^2}  
		\right] .
	\end{split}
\end{equation}
%
In \autoref{fig:FigureSM_esMod_Vim} these two contributions are plotted as a function of the position of the negative charge 
when the positive charge is at $z_0/d = 0.2$. The top (bottom) panel shows the energy as a function of the $z_d$-coordinate ($x_d$-coordinate) when $x_d = y_d = 0$ ($y_d = z_d = 0$).

In the limit, where the size of the dipole is small compared to $d$, the lowest corrections to the ground state energy will be given by the second-order expansion of $V_{\rm im}$, which is given by 
\begin{equation} \label{eq:SM1_Vimexpansion}
	\begin{split}
		V^{(2)}_{\rm im} = - \frac{q^2}{4 \pi \epsilon_0d^3}  & \left\{\frac{x_d^2 + y_d^2}{32} \left[ \frac{2}{(z_0/d)^3} -\Psi^{(2)} \left(1-\frac{z_0}{d}\right) \right. \right.
		\\  & \left. \left. \hspace{17mm} - \Psi^{(2)} \left(1+\frac{z_0}{d}\right) - 4 Z(3)
		\right]
		\right. 
		\\
	&	\left. \hspace{7mm} +  \frac{z_d^2}{16} \left[ \frac{2}{(z_0/d)^3} -\Psi^{(2)} \left(1-\frac{z_0}{d}\right) \right. \right. \\ & \left.\left.  \hspace{17mm} - \Psi^{(2)} \left(1+\frac{z_0}{d}\right) + 4 Z(3)
		\right]
		\right\}.
	\end{split}
\end{equation}
Here, $Z$ is the Riemann zeta function and the polygamma function $\Psi^{(2)}(z) $ is defined by 
\begin{equation}
\Psi^{(n)}(z) = \frac{\dd^{n+1} }{\dd z^{n+1}}  \log \Gamma(z), 
\end{equation}
where $\Gamma (z)$ is the Euler gamma function. This correction $V^{(2)}_{\rm im}$ is plotted against the total contribution $V_{\rm im}$ in \autoref{fig:FigureSM_esMod_Vim} (dotted red and dashed purple lines, respectively).

\section{Ground state energy for the $LC$-resonator setup}
\label{sec:SMmodEnergyLC}

In this section we present the derivation of the energy shift experienced by the ground state of the matter system in the $LC$-resonator setup described in \autoref{sec:SMcavityQEDHamiltonianLC}. This shift can be decomposed into three different contributions,
\begin{equation}
\Delta E_{\rm GS} = \Delta E_{\rm im} + \Delta E_{\bA} + \Delta E_{\rm cav} \ ,
\end{equation}
which come from electrostatic corrections, the coupling to the transverse modes within the cavity, and the coupling to the $LC$ resonance, respectively. With the purpose of determining the precise value of these corrections, we will consider the particular case of an isotropic harmonic dipole of mass $m$ and frequency $\omega_0$.  The wavefunction describing the ground state for the three-dimensional harmonic oscillator reads
\begin{equation} \label{eq:SMes_3DharmoscGS}
	\psi_{\rm GS} ( \br_d) = \left(\frac{1}{2 \pi a_0^2}\right)^{3/4} \e^{-\br_d^2/(4a_0^2)}   \ ,
\end{equation}
where $a_0 \equiv \sqrt{\hbar /(2 m \omega_0)}$ is the characteristic length of the dipole. Nonetheless, up to numerical prefactors, our predictions are valid for other cavity QED systems with a similar transition frequency.

\subsection{Electrostatic effects}
\begin{figure}[bt!] 
	\centering
	\includegraphics[width=\linewidth]{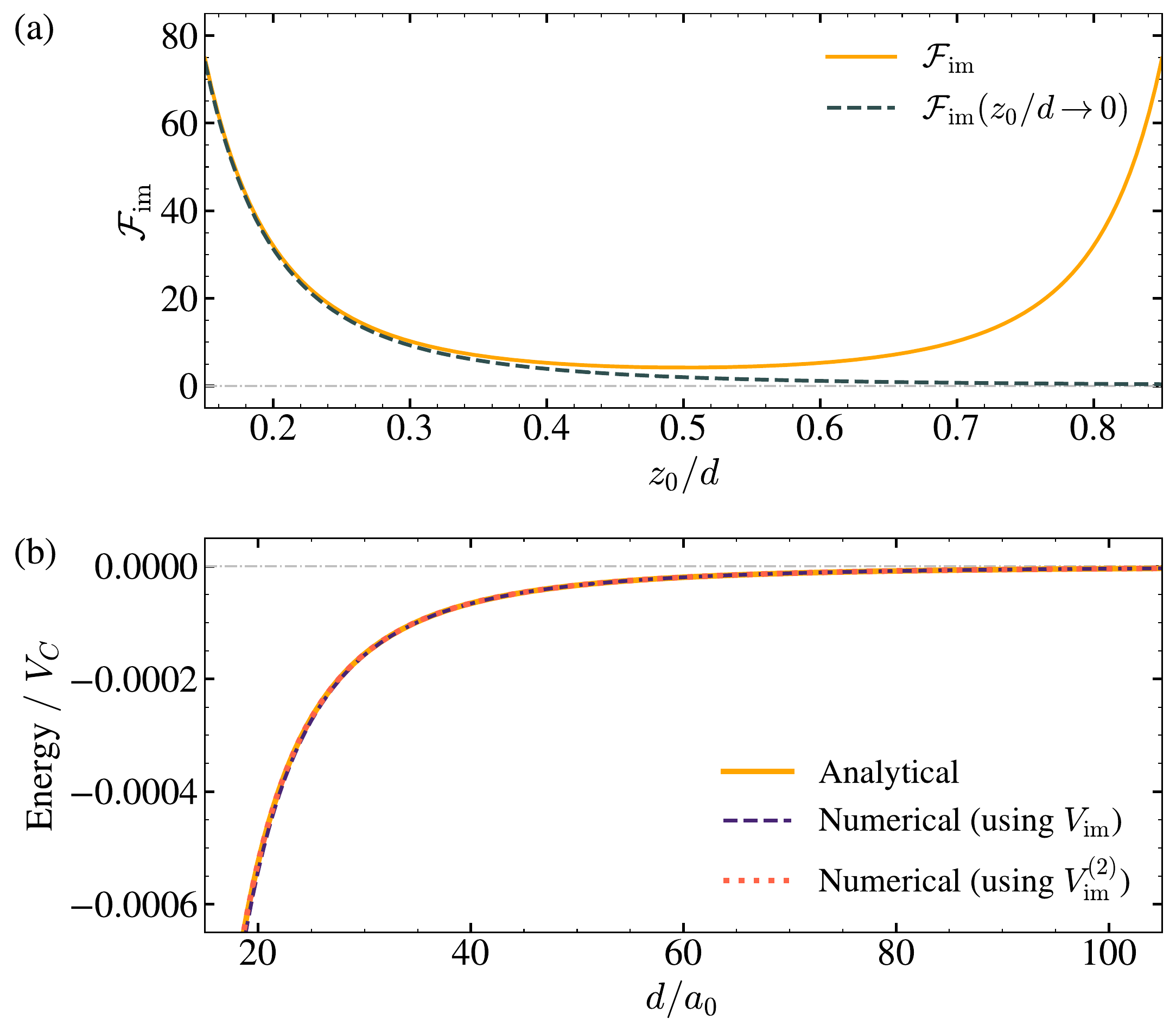}
	\caption{(a) Numerical prefactor $\mathcal{F}_{\rm im}$ governing the modification of the ground state energy due to electrostatic effects as a function of the relative position of the positive charge, $z_0/d$. The continuous line plots the complete expression (\autoref{eq:SMes_Fimdef}) while the dashed line plots its approximation in the limit $z_0/d \rightarrow 0$ (\autoref{eq:SMes_FimdefSerie}). (b) Ground state energy shift due to electrostatic effects as a function of the plate separation $d$.  The analytical expression for the energy shift (\autoref{eq:SMes_deltaEimexpAna}) is plotted in continuous yellow line against its numerical evaluation from \autoref{eq:SMes_deltaEimexp} using both $V_{\rm im}$ (dashed purple line) and $V_{\rm im}^{(2)}$ (dotted red line).}
	\label{fig:FigureSM_esMod_Fim}
\end{figure}

Due to the presence of induced charges on the cavity mirrors, the electrostatic energy presents a correction $V_{\rm im}$ to that corresponding to free space, namely, $V = V_{\rm free} + V_{\rm im}$. To lowest order in perturbation theory this induces a shift of the dipole's ground state energy of
\begin{equation} \label{eq:SMes_deltaEimexp}
		\Delta E_{\rm im} = \langle \psi_{\rm GS} | V_{\rm im} | \psi_{\rm GS} \rangle.
\end{equation}
In general, this expression can be evaluated numerically by averaging the full expression for $V_{\rm im}$ in \autoref{eq:SM1_Vimexp1} over the ground state wavefunction. However, for most situations of interest, a good estimate of this shift  can already be obtained by taking only the second-order expansion, $V_{\rm im}^{(2)}$ in \autoref{eq:SM1_Vimexpansion} into account, in which case  
\begin{equation} \label{eq:SMes_deltaEimexpAna}
		\Delta E_{\rm im} \simeq \langle \psi_{\rm GS} | V_{\rm im}^{(2)}| \psi_{\rm GS} \rangle
		= - V_C  \frac{a_0^3}{d^3} \mathcal{F}_{\rm im}.
\end{equation}
Here, $V_C = q^2 /( 4 \pi \epsilon_0 a_0)$ and we have defined the numerical prefactor 
\begin{equation} \label{eq:SMes_Fimdef}
	\begin{split}
		\mathcal{F}_{\rm im}= \frac{1}{8}    \left[ \frac{2}{(z_0/d)^3} -\Psi^{(2)} \left(1-\frac{z_0}{d}\right)- \Psi^{(2)} \left(1+\frac{z_0}{d}\right)
		\right]	 .
	\end{split}
\end{equation}
In the limit $z_0/d \rightarrow 0$, this factor can be approximated by
\begin{align} 
		\mathcal{F}_{\rm im} & = \frac{1}{4(z_0/d)^3} - \frac{1}{4} \Psi^{(2)} \left(1 \right) + o(z_0/d) 
		\\ \label{eq:SMes_FimdefSerie} & 		\approx \frac{1}{4(z_0/d)^3} 
		\ ,
\end{align}
thus reproducing the van der Waals interaction for a dipole close to a single plate,
\begin{equation} 
	\Delta E_{\rm im}^{(\rm vdW)} = - \frac{q^2} {4 \pi \epsilon_0 } \frac{a_0^2}{4z_0^3}  \ .
\end{equation}
In \autoref{fig:FigureSM_esMod_Fim} we plot the factor $\mathcal{F}_{\rm im}$ as a function of the position of the positive charge, $z_0/d$, as well as its approximation in the single-plate limit.%

\subsection{Transverse modes}

The ground state energy also becomes modified due to the coupling to the transverse electromagnetic field confined within the plates, where the interaction Hamiltonian is given by
\begin{equation}  \label{eq:SMLamb_intHamiltonian}
	H_{\perp}=  \frac{q}{m} \bA \cdot \bp + \frac{q^2}{2m} \bA^2.
\end{equation}
For the potential operator $\bA$ we use the mode expansion in \autoref{eq:Aopexpansion}.
Given the symmetry of our setup, the mode functions ${\bu}_\lambda ({\bx})$ in \autoref{eq:SM3_modesModeFunctionsD3} can be rewritten in terms of the two-dimensional vectors $\boldsymbol \rho$, $\bk_{\parallel}$, and  ${\bepsl}_\parallel$ giving the components parallel to the $xy$-plane of the corresponding three-dimensional vectors, namely, $\br : (\brho, z)$, $\bk : (\bk_{\parallel}, k_z)$, and $\bepsl : ({\bepsl} _\parallel, {\epsl}_z)$, yielding
%
\begin{equation} \label{eq:SM3_modesModeFunctionsD4}
	\bul (\bx) = \sqrt{\frac{2}{V (1 + \delta_{n0})}} \ \ee^{\iii \bk_\parallel \cdot \brho}
	\begin{pmatrix}
		\ii {\bepsl} _\parallel \sin (k_z z)	\\
		{\epsl} _z \cos (k_z z)
	\end{pmatrix}  .
\end{equation}
%
From \autoref{eq:SM1_modesBounCond}, we can express $k_{\parallel} = |\bk_\parallel |$ and  $k = |\bk|$ in terms of the indices $\kappa = \sqrt{n_x^2 + n_y^2} $ and $\nu = \sqrt{({4}/{x^2}) \kappa^2 + n^2}$,
\begin{align}
	\label{eq:SMLamb_modes_kparalleliskappa} 
	k _ \parallel &  = \sqrt{k_x^2 + k_y^2} =  \frac{2 \pi}{l} \kappa =  \frac{\pi}{d} \frac{2 \kappa}{x} \ ,
	\\            \label{eq:SMLamb_modes_kisnu} 
	k   & = \sqrt{k _ \parallel^2 + k_z^2 } = \frac{\pi}{d} \nu
	\ ,
\end{align}
where the length $l$ is given in terms of the distance $d$ between the plates through the ratio parameter $x$, that is, $l = xd$. Likewise, we can define the wavevector $k_0$ and the dimensionless number $\nu_0$ associated with the transition within the molecular dipole,
%
\begin{equation} \label{eq:SMLamb_modes_kmisnum}
	\omega_{0} = c k_0 = \frac{\pi c }{d} \nu_0 \ .
\end{equation}

%

\subsubsection{Perturbation theory and renormalization}

Using perturbation theory up to second order in $q$, the correction to the ground state energy presents two contributions,
\begin{equation} 
	 \Delta \tilde E_{\bA} = \Delta  E_{\bA^2} +  \Delta  E_{\bA\cdot \bp}  \ ,
\end{equation}
coming from each of the terms in the interaction Hamiltonian (\autoref{eq:SMLamb_intHamiltonian}), with
\begin{align}
	\label{eq:SMLamb_DeltaEA2cont0}
		& \Delta  E_{\bA^2} =  \frac{q^2 \hbar}{2mc \epsilon_0 V} {\sum_{\bk}}'
		 \frac{1}{k}
		\left[ 1 -  \frac{k_z^2}{k^2} \cos (2 k_z z_0)\right]  ,
		\\  \nonumber
		&\Delta  E_{\bA \cdot \bp} = -\frac{q^2}{m^2 c^2 \epsilon_0 V} \sum_{j>0}  {\sum_{\bk}}'
		\frac{1}{ k ( k +  k_j)} \\ \label{eq:SMLamb_DeltaEApcont0}
		& \left[\frac{1}{2}\left(1 + \frac{k_z^2}{k^2}\right) \sin^2(k_z z_0) p_{j\parallel}^2  + \left(1 - \frac{k_z^2}{k^2}\right) \cos^2(k_z z_0) p_{jz}^2 \right]   .
\end{align}
The prime in the summation over $\bk$ recalls the extra factor 1/2 associated with $n=0$ (coming from the renormalization of the mode functions in \autoref{eq:SM3_modesModeFunctionsD4}) and the summation over polarizations has been already performed using that
\begin{align}
\sum_\sigma |\varepsilon_{{\bk} \sigma z}|^2  = 1 - \frac{k_z^2}{k^2} \ , \hspace{5mm}
\sum_\sigma |\varepsilon_{{\bk} \sigma \parallel}|^2  = 1 + \frac{k_z^2}{k^2} \ .
\end{align}
Besides, we have defined the matrix elements $p_{j\parallel}^2 \equiv |\langle j | p_x | 0\rangle|^2 + |\langle 1 |  p_y | 0 \rangle|^2$ and $p_{jz}^2 \equiv | \langle j | p_z | 0 \rangle |^2$, where the index $j$ labels all possible excited states of the matter system (characterized by the transition frequency $\omega_j  = ck_j$ to the ground state). In the case of the harmonic oscillator, the first excited state is the only one that contributes, and $p_{z}^2 = p_{\parallel}^2 /2  = \hbar m \omega_m /2$. Therefore, \autoref{eq:SMLamb_DeltaEApcont0} simplifies to
\begin{equation}
	\begin{split} \label{eq:SMLamb_DeltaEApcont1}
		\Delta E_{\bA \cdot \bp} = -\frac{q^2 \hbar}{2 m c \epsilon_0 V}   {\sum_{\bk}}'
		\frac{k_m}{ k ( k +  k_m)} 
		 \left[1 - \frac{k_z^2}{k^2} \cos (2k_z z_0) \right].   
	\end{split}
\end{equation}
%
When the plates have infinite extension ($l \rightarrow \infty$), we can transform the sum over the parallel component of the wavevector into an integral, yielding
\begin{equation} \label{eq:SMLamb_DeltaEA2cont2}
\begin{split}
	\Delta E_{\bA^2} & =   \frac{q^2 \hbar}{mc \epsilon_0 l^2} \int_0^\infty  \dd {\kappa} \ \kappa   {\sum_{n}}'   \frac{ 1}{\nu}\left[ 1 -  \frac{n^2}{\nu^2}  \cos \left(\frac{2\pi n z_0}{d}\right) \right]\\
	& = \frac{q^2 \hbar}{ 4mc \epsilon_0 d^2} {\sum_{n}}' \int_n^\infty  \dd \nu\, \left[ 1 -  \frac{n^2}{\nu^2}  \cos \left(\frac{2\pi n z_0}{d}\right) \right] ,
\end{split}
\end{equation}
where in the second line we changed the integration variable from $\kappa$ to $\nu$. Similarly, 
\begin{equation}\label{eq:SMLamb_DeltaEApcont2}
\begin{split}
	\Delta E_{\bA \cdot \bp}  = -\frac{q^2 \hbar }{4m c \epsilon_0 d^2}  {\sum_{n}}'   \int_n^\infty \dd \nu  &\frac{\nu_0}{\nu +  \nu_0}\\
	&\left[ 1  - \frac{n^2}{\nu^2}   \cos \left(\frac{2\pi n z_0}{d}\right) \right].
\end{split}
\end{equation}
%
%

The sums in \autoref{eq:SMLamb_DeltaEA2cont2} and \autoref{eq:SMLamb_DeltaEApcont2} are performed over an infinite number of modes, which would lead to an infinite value of the energy shift. To obtain a finite expression for the energy shift, we must keep in mind that $\Delta E_{\bA}$ is defined relative to the ground state energy in free space,
\begin{equation}
\Delta E_{\bA}= \Delta \tilde E_{\bA} - \left. \Delta \tilde E_{\bA}\right|_{\rm free}.
\end{equation}
We can obtain the free-space contribution from \autoref{eq:SMLamb_DeltaEA2cont2} and \autoref{eq:SMLamb_DeltaEApcont2} above by replacing the discrete sum over $n$ by an integral and ignoring the terms that depend explicitly on $z_0$. As a result, the final expression for the energy shift $\Delta E_{\bA}$ can be written as
\begin{equation}\label{eq:SMLamb_DeltaEA_final}
\begin{split}
\Delta E_{\bA} = \alpha\hbar \omega_m \left(\frac{q a_0}{e d}\right)^2 \mathcal{F}_{\bA},
\end{split}
\end{equation}%
where
\begin{equation} \label{eq:SMLamb_prefactorFAdef}
\mathcal{F}_{\bA}=2\pi\left[ \funAAg + \funAAf (z_0/d) - \funApg (\nu_0) - \funApf (\nu_0, z_0/d)\right]
\end{equation}%
is a dimensionless function. Here we have defined the two quantities
\begin{align} \label{eq:SMLamb_g1def}
	{\funAAg}  & =   {\sum_{n}}'  \int_{n}^\infty   \dd \nu - \int_0^\infty \dd n \  \int_{n}^\infty   \dd \nu  \ , \\
	\label{eq:SMLamb_f1def}
	{\funAAf} (z_0/d) & = -  {\sum_{n}}'    n^2 \cos \left(\frac{2\pi n z_0}{d}\right)   \int_{n}^\infty  \frac{\dd {\nu}   }{\nu^2} \ ,
\end{align}
which arise from the evaluation of the $\bA^2$ contribution. The other two functions 
\begin{align} \label{eq:SMLamb_gmdef}
	\funApg (\nu_0) & =   
	 {\sum_{n}}'  \int_n^\infty 
	\frac{\nu_0  }{ \nu +  \nu_0} \dd \nu
	\\
	&- \int_0^\infty  \dd n  \int_n^\infty 
	\frac{\nu_0 }{\nu +  \nu_0}  \dd \nu
\end{align}
and
\begin{align} 
	\label{eq:SMLamb_fmdef}
	\funApf (\nu_0, z_0/d) &  = -   {\sum_{n}}'  \cos \left(\frac{2\pi n z_0}{d}\right)
	\int_n^\infty \frac{n^2   \nu_0 }{\nu^2 (\nu +  \nu_0)}  \dd \nu
\end{align}
arise from the evaluation of the $\bA \cdot \bp$ term and depend on the dipole frequency $\nu_0$.

\subsubsection{Evaluation of the individual sums and integrals}
The expressions in \autoref{eq:SMLamb_g1def}-\autoref{eq:SMLamb_fmdef} have previously been derived in Ref.~\cite{Barton1970_SM}, where the author then introduced a sharp frequency cutoff $\Lambda$ to evaluate the otherwise diverging sums and integrals. However, working with a sharp cutoff is somewhat ambiguous since some of the expressions do not converge for $\Lambda\rightarrow \infty$ and an additional ad-hoc averaging procedure must be performed in order to get meaningful results. In the following we use a combined analytic and numerical approach to evaluate \autoref{eq:SMLamb_g1def}-\autoref{eq:SMLamb_fmdef} also for a smooth and therefore more physical cutoff function. Importantly, this analysis shows that all results are independent of the details of the cutoff function and also that convergence is reached for rather modest values of the cutoff. We further provide additional analytic results for the limit of strong confinement, $\nu_0\rightarrow0$, which is the regime most relevant for the current study.

We start with the evaluation of the function $\funAAg$ in \autoref{eq:SMLamb_g1def}, which is the difference between a diverging sum and a diverging integral. To obtain a finite result, we need to introduce a cutoff scale $\Lambda$ at high frequency for both the sum and the integral~\cite{Casimir1984_SM}. We can use either a sharp cutoff at $\nu = \Lambda$ or multiply the integrand by a smooth cutoff function $h_c (\nu / \Lambda)$. In particular, we consider a cutoff function that satisfies $h_c (0) = 1$, $h_c'(0)\simeq 0$, and $h_c(\nu/ \Lambda \gg 1) \rightarrow 0$ sufficiently fast.
In that case, \autoref{eq:SMLamb_g1def} transforms to
\begin{equation}
	\funAAg  = - \frac{1}{2} F(0, \nu/\Lambda) +  {\sum_{n=0}^\infty} F(n, \nu/\Lambda) - \int_0^\infty \dd n \ F(n, \nu/\Lambda)  \ , 
\end{equation}
with $F(n, \nu/\Lambda) = \int_{n}^\infty h_c(\nu/\Lambda) \ \dd \nu$ and where we have already taken into account the extra factor associated with the summand $n=0$.
The difference between the sum and the integral can be evaluated using the Euler--Maclaurin formula and we obtain
\begin{equation}
	\begin{split}
	\funAAg  =  -  \frac{B_2}{2}  F'(0, \nu/\Lambda)+ o[F'''(0, \nu/\Lambda)] \ , 
		\end{split}
\end{equation}
where $B_2=1/6$ is the second Bernoulli number and the apostrophe indicates derivative with respect to $n$. We have assumed that $F(n, \nu/\Lambda)$ and all its derivatives go to zero for $n \rightarrow \infty$. Further, the higher-order derivatives of $F$ decrease with increasing $\Lambda$ and can therefore be neglected. Since $F'(0, \nu/\Lambda) = - h_c(0) = -1$, we finally have that
\begin{equation} \label{eq:SMLamb_g1ana}
	\funAAg  = \frac{1}{12},  
\end{equation}
regardless of the precise details of the cutoff function. 

\begin{figure}[t]	
	\centering
	\includegraphics[width=\linewidth]{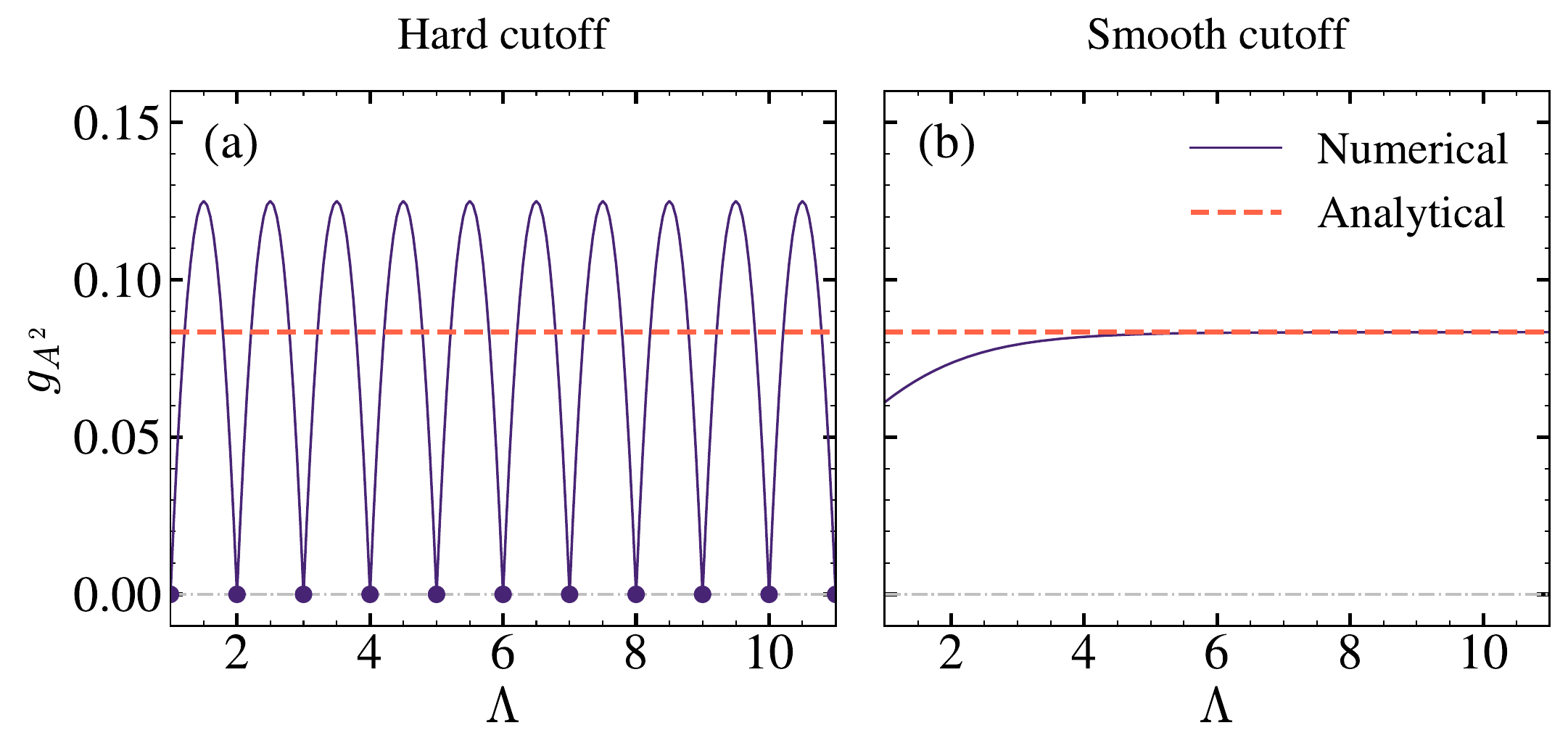}
	\caption{Numerical evaluation of $\funAAg$ (\autoref{eq:SMLamb_g1def}) as a function of the cutoff $\Lambda$ when using: (a) a sharp cutoff at $\nu = \Lambda$, (b) the smooth cutoff function $h_c (\nu/\Lambda) = 1-1/(1+\e^{-\nu+\Lambda})$. In the  evaluation of the sum in \autoref{eq:SMLamb_g1def} for a hard cutoff, the value of $\Lambda$ has been rounded down to the nearest integer. Purple dots in (a) highlight the points where the cutoff takes an integer value. The dashed red line indicates the analytical result (\autoref{eq:SMLamb_g1ana}).}
	\label{fig:SMLamb_g1}
\end{figure}

Note that the same result has been obtained in Ref.~\cite{Barton1970_SM} by introducing a sharp cutoff, i.e., $h_c(\nu/\Lambda)=\Theta(1-\nu/\Lambda)$, where $\Theta$ is the unit step function. However, as shown in~\autoref{fig:SMLamb_g1} (a) this creates an ambiguity, since in this case the value of $\funAAg$ varies periodically with $\Lambda$ and $\funAAg=1/12$ is only obtained on average. In contrast, for the smooth cutoff assumed in~\autoref{fig:SMLamb_g1} (b), the value of $\funAAg$ converges toward the analytic prediction. This plot also shows that this convergence happens already for very small values of $\Lambda\approx 5$. 

Now we turn to the computation of the term that depends on the particle position, $\funAAf$ (\autoref{eq:SMLamb_f1def}). By solving the integral over $\nu$, we obtain
\begin{equation} \label{eq:SMLamb_f1numsum0}
	\funAAf (\zeta)  = - \sum_{n=1}^\infty  (-1)^n  n \cos (n \zeta)   ,
\end{equation}
where we have expressed the $z$-position of the dipole in terms of the parameter $\zeta$,
\begin{equation} \label{eq:SMLamb_zetaDef}
	z_0 = \frac{d}{2 \pi}(\pi + \zeta) ,
\end{equation}
with $\zeta \in [-\pi, \pi]$.  Taking into account that 
$n^2 \cos (n \zeta)  = - {\partial^2 \cos (n \zeta)}/{\partial \zeta^2}$, we obtain
\begin{equation} 
\begin{split}
	\funAAf (\zeta) &=\frac{\partial^2 }{\partial \zeta^2} \sum_{n=1}^\infty  \frac{(-1)^n}{n}  \cos (n \zeta) \\
	 &= \frac{1}{2}  \frac{\partial^2 }{\partial \zeta^2} \ln \left[\frac{1}{2+2\cos(\zeta)}\right]
\end{split}
\end{equation}
and, finally,~\cite{Barton1970_SM}
\begin{equation} \label{eq:SMLamb_f1ana}
	\funAAf (z_0/d) = \frac{1}{4 \sin^2 (\pi z_0 / d)} .
\end{equation}

Note that in this derivation we have exchanged the derivative with summation, which implicitly assumed a convergence of the sum. To understand the validity of this approach in the presence of a finite cutoff, we can also evaluate $\funAAf$ numerically from \autoref{eq:SMLamb_f1def}. As above, a high-frequency limit can be introduced via a smooth cutoff function $h_c(\nu/\Lambda)$ that multiplies the integrand or using a sharp cutoff at $\nu = \Lambda$. The latter yields the following expression for $\funAAf$ in the form of a sum,
\begin{equation} \label{eq:SMLamb_f1numsum}
	\funAAf (z_0/d)  = - \sum_{n=1}^{[\Lambda]}   n^2 \cos (2 \pi n z_0/d) \left( \frac{1}{n} - \frac{1}{\Lambda} \right)  ,
\end{equation}
where $[\Lambda]$ denotes the integer part of $\Lambda$.

\begin{figure}[tb]	
	\centering
	\includegraphics[width=0.98\linewidth]{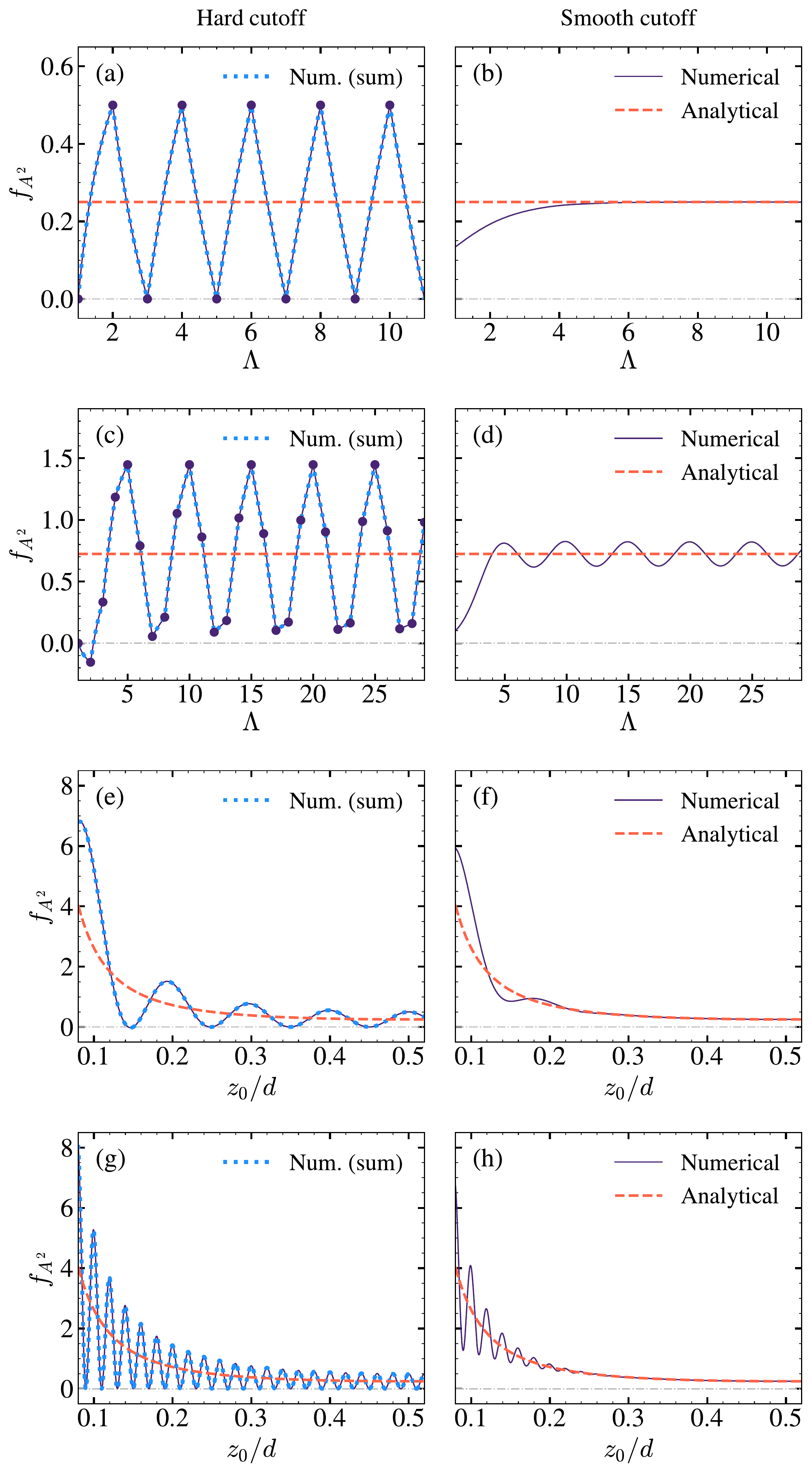}
	\caption{Numerical evaluation of $\funAAf$ from \autoref{eq:SMLamb_f1def} (continuous dark blue line) and \autoref{eq:SMLamb_f1numsum} (dotted light blue line) using either a sharp cutoff at $\nu = \Lambda$ (left panels) or the smooth cutoff function $h_c (\nu/\Lambda) = 1-1/(1+\e^{-\nu+\Lambda})$ (right panels). This magnitude is plotted versus $\Lambda$ when $z_0/d = 0.5$ (a-b) and $z_0/d = 0.2$ (c-d), and versus the position of the particle $z_0/d$ when $\Lambda = 10$ (e-f) and $\Lambda = 50$ (g-h). 	The value of $\Lambda$ in the sum appearing in \autoref{eq:SMLamb_f1def} has been rounded down to the nearest integer when using a hard cutoff, and the purple dots in (a) and (c) indicate the points where the cutoff takes an integer value. Dashed red lines stand for the analytical result  (\autoref{eq:SMLamb_f1ana}).}
	\label{fig:SMLamb_f1}
\end{figure}

In \autoref{fig:SMLamb_f1} we plot the numerical evaluation of $\funAAf$ using either a sharp (left panels) or a smooth (right panels) cutoff from both \autoref{eq:SMLamb_f1def} (continuous dark blue line) and \autoref{eq:SMLamb_f1numsum} (dotted light blue line). The analytical result (\autoref{eq:SMLamb_f1ana}) is displayed with a dashed red line for comparison. The value of $\funAAf$ is plotted as a function of the cutoff $\Lambda$ for $z_0/d = 0.5$ in panels (a-b) and $z_0/d = 0.2$ in panels (c-d).  For $z_0/d=0.5$ we find a similar behavior as in the evaluation of $\funAAg$ above and in the case of a smooth cutoff, a quick convergence to the analytically predicted value. For $z_0$ close to the metallic plates, neither a sharp cutoff nor a smooth cutoff function lead to converging results and the value of $\funAAf(\Lambda)$ oscillates around its analytic prediction even for very large values of $\Lambda$. Instead, as $\Lambda$ increases, we observe faster and faster oscillations of $\funAAf$ as a function of $z_0$. However, as shown in \autoref{fig:SMLamb_f1} (e-h), these oscillation are considerably less pronounced in the case of a smooth cutoff, where the analytic results are still recovered for a moderate $\Lambda$, expect in a region very close to the boundaries. 

\begin{figure}[tb]	
	\centering
	\includegraphics[width=\linewidth]{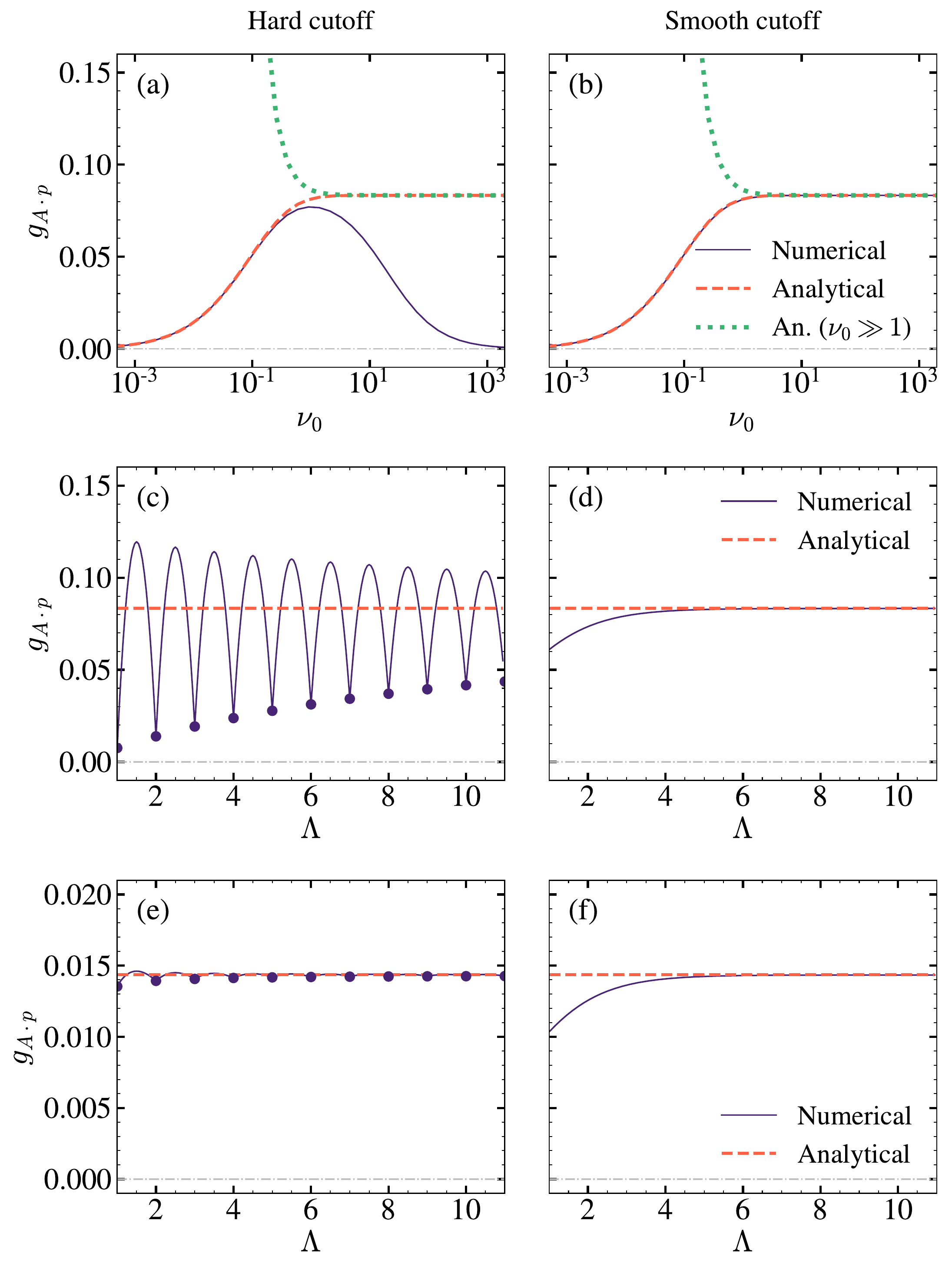}
	\caption{Numerical evaluation of $\funApg$ from \autoref{eq:SMLamb_gmdef} (continuous blue line) using either a hard cutoff $\Lambda$ (left panels) or the smooth cutoff function $h_c (\nu/\Lambda) = 1-1/(1+\e^{-\nu+\Lambda})$ (right panels). The analytical expression for $\funApg$ given in \autoref{eq:SMLamb_g0analow} is plotted with a dashed red line, while the dotted green line represents its analytical approximation in the high-frequency limit (\autoref{eq:SMLamb_g0anahigh}). The value of $\funApg$ is plotted either versus the factor $\nu_0$ for $\Lambda =10$ (a-b) or versus the cutoff $\Lambda$ for $\nu_0 = 10$ (c-d) and $\nu_0=0.01$ (e-f).
	In the  evaluation of the sum in \autoref{eq:SMLamb_gmdef} when using a hard cutoff, the value of $\Lambda$ has been rounded down to the nearest integer. Purple dots in panels (c) and (e) mark the points where the cutoff takes an integer value.	
	}
	\label{fig:SMLamb_gm}
\end{figure}

In a next step we consider the contributions coming from the $\bA \cdot \bp$ term of the interaction Hamiltonian, that is, $\funApg$ and $\funApf$.
Analogously to the computation of $\funAAg$, we can also 
use a cutoff function $h_c (\nu / \Lambda)$ that prevents the integral over $\nu$ in $\funApg$ (\autoref{eq:SMLamb_gmdef}) from diverging. In particular, if we define  $G(n, \nu/\Lambda) = \int_n^\infty [{\nu_0 }/({\nu +  \nu_0})]  h_c (\nu/\Lambda)  \dd \nu $, the term $\funApg$ can be rewritten as follows,
\begin{equation} 
	\funApg = -\frac{1}{2}  G(0, \nu/\Lambda) + 
	{\sum_{n=0}^\infty}  G(n, \nu/\Lambda)
	- \int_0^\infty  \dd n \  G(n, \nu/\Lambda)
	\ , 
\end{equation}
where we have made explicit the different factor for the term with $n=0$ in the summation. Using the Euler--Maclaurin formula, we have that
\begin{equation}\label{eq:SMLamb_gm_highfreq_expand}
	\begin{split}
		\funApg  = -  \frac{B_2}{2}  G'(0, \nu/\Lambda) -\frac{B_4}{24}[G'''(0, \nu/\Lambda)] +\dots
	\end{split}
\end{equation}
where $B_4=-1/30$. By using the same arguments as for the calculation of $\funAAg$, we obtain that
\begin{equation} \label{eq:SMLamb_g0anahigh}
\funApg \simeq \frac{1}{12} + \frac{1}{360\nu_0^2} +\dots
\end{equation}
which is also the result obtained in Ref.~\cite{Barton1970_SM}. However, the higher-order derivatives in the Euler--Maclaurin formula now scale as
\begin{equation}
	G^{(k)} (0, \nu/\Lambda) \sim \frac{1}{\nu_0^{(k-1)}} 
\end{equation}
for small $\nu_0$. Therefore, this expansion does not converge in the limit $\nu_0 \rightarrow 0$. This can be observed in \autoref{fig:SMLamb_gm}(a-b), where the dotted green line indicates this analytical value of $\funApg$ as a function of $\nu_0$. This prediction agrees with the numerical result obtained from the direct evaluation of \autoref{eq:SMLamb_gmdef} with a sharp and a smooth cutoff for large values of $\nu_0$, but exhibits a very different behavior when $\nu_0\ll 1$. 

To obtain a result for $\funApg$ that is also valid at low-frequencies, we write the full expression for $\funApg$ as 
\begin{equation} \label{eq:SMLamb_gmExp1}
\begin{split} 
	\funApg  =& \frac{1}{2}  G(0, \Lambda) - \int_0^1 \dd n \  G(n, \Lambda) \\
		&+{\sum_{n=1}^\infty}  G(n, \Lambda)
		- \int_1^\infty \dd n \  G(n, \Lambda). 
\end{split}
\end{equation}
We now choose a sharp cutoff function, $h_c (\nu/\Lambda) = \Theta(1 - \nu/\Lambda)$, in which case
\begin{equation}
G (n, \Lambda) = \nu_0 \ln \left( \frac{\nu_0 + \Lambda}{\nu_0 + n}\right)\Theta(\Lambda-n) \ ,
\end{equation}
and we can evaluate the first line in \autoref{eq:SMLamb_gmExp1} explicitly. For the second line we use the Euler-Maclaurin formula and obtain
\begin{equation}  
	\begin{split}
	\funApg  = & \frac{1}{2}  G(0,\Lambda) + \frac{1}{2}  G(1, \Lambda)
	- \int_0^1  \dd n \  G(n, \Lambda)
	\\
	& 	-\frac{B_2}{2!}  G'(1, \Lambda) - \frac{B_4}{4!} G'''(1, \Lambda)  + \ldots
	\end{split}
\end{equation}
In this expansion the derivatives in the Euler-Maclaurin formula are evaluated at $n=1$, and therefore give finite values for $\nu_0\rightarrow 0$. To a first approximation, we take only the first term of this expansion, $-B_2  G'(1, \Lambda)/2=\nu_0/(12(1+\nu_0))$, into account, which results in 
\begin{equation} \label{eq:SMLamb_g0analow}
	\begin{split}
	\funApg  \simeq \nu_0 \left[\left(\nu_0 + \frac{1}{2}\right) \ln \left(\frac{\nu_0+1}{\nu_0}\right)  + \frac{1}{12(1+\nu_0)}-1\right].
	\end{split}
\end{equation}
In \autoref{fig:SMLamb_gm} (a-d), the dashed red line shows this expression as a function of $\nu_0$, showing a perfect agreement with the exact numerics for all frequencies. The dependence of the numerical result with the cutoff value is explored in \autoref{fig:SMLamb_gm} (c-d) for $\nu_0=10$ and in \autoref{fig:SMLamb_gm} (e-f) for $\nu_0=0.01$. 

Finally, we compute the term that depends on the position of the dipole, $\funApf$ (\autoref{eq:SMLamb_fmdef}). Solving the integral over $\nu$, we have
\begin{equation} \label{eq:SMLamb_f0numsum}
	\funApf (\zeta)  = - \sum_{n=1}^\infty (-1)^n n^2 \cos (n\zeta)   
	\left[  \frac{1}{n } + \frac{1}{\nu_0} \ln \left(\frac{n}{n+\nu_0}\right)
	\right],
\end{equation}
where, again, we have expressed the position of the dipole in terms of the parameter $\zeta$ (\autoref{eq:SMLamb_zetaDef}). We evaluate this contribution in the low-frequency and in the high-frequency limit separately. For $\nu_0\gg1$ we follow the analysis of Ref.~\cite{Barton1970_SM} and write this function as $\funApf (\zeta)  = \funAAf(\zeta) +\funtildeApf (\zeta)$, where $\funAAf$ has already been evaluated in \autoref{eq:SMLamb_f1ana} and 
\begin{align} 
		\funtildeApf (\zeta) & = - \frac{1}{\nu_0} \sum_{n=1}^\infty (-1)^n n^2 \cos (n\zeta)   
		\ln \left(\frac{n}{n+\nu_0}\right) 
		\\  \label{eq:SMLamb_f0numint}
		& = - \frac{1}{\nu_0} \int_0^\infty \dd u \  u^2 \ \frac{\cosh(u \zeta)}{\sinh(u \pi)} \arctan \left(\frac{\nu_0}{u}\right).
\end{align}
This last identity can be obtained using contour integration~\cite{Barton1970_SM}. In the limit $\nu_0/\kappa \rightarrow \infty$, we can now expand $\arctan (\nu_0/u)$ in powers of  $1/\nu_0$, which yields
\begin{align}\label{eq:SMLamb_f0anagg}
	\funtildeApf (\zeta)  =
	& + \frac{\pi}{2\nu_0} \int_0^\infty \dd u \  u^2 \ \frac{\cosh(u \zeta)}{\sinh(u \pi)}
	\\ \nonumber
	& + \frac{1}{\nu_0^2} \int_0^\infty \dd u \  u^3 \ \frac{\cosh(u \zeta)}{\sinh(u \pi)}  + \ldots
	\\
	=	& + \frac{\pi}{2\nu_0} \frac{1}{(2\pi)^3} 
	\left[ \Psi^{(2)} \left(\frac{z_0}{d}  \right) + \Psi^{(2)} \left(1- \frac{z_0}{d}   \right) \right]  
	 \nonumber \\ \nonumber
	& + \frac{1}{\nu_0^2} \frac{3 - 2 \sin^2 (\pi z_0/d)}{8 \sin^4(\pi z_0/d)}  + \ldots 
\end{align}
We can further apply the recurrence relation satisfied by the polygamma functions, 
$\Psi^{(m)} (z+1) = \Psi^{(m)} (z) + (-1)^m m!/z^{m+1}$ to write 
\begin{equation}
\Psi^{(2)} \left(\frac{z_0}{d}  \right)= \Psi^{(2)} \left(1+\frac{z_0}{d}  \right)- \frac{2}{(z_0/d)^3}.
\end{equation} 
Overall, we obtain in the high frequency limit
\begin{equation}
\funApf (z_0/d)= \frac{1}{4\sin^2(\pi z_0/d)}- \frac{\mathcal{F}_{\rm im}(z_0/d)}{2\pi^2 \nu_0}  + o ( \nu_0^{-2}).
\end{equation}
The appearance of the function $\mathcal{F}_{\rm im}(z_0/d)$, which we have already encountered in the derivation of the electrostatic corrections (\autoref{eq:SMes_Fimdef}), is of course not coincidental. Indeed, this term cancels the electrostatic contribution exactly~\cite{Barton1970_SM}, since the limit of large distances and high frequencies, $\nu_0$, the interaction with the walls must be fully retarded.

\begin{figure}[th]	
	\centering
	\includegraphics[width=\linewidth]{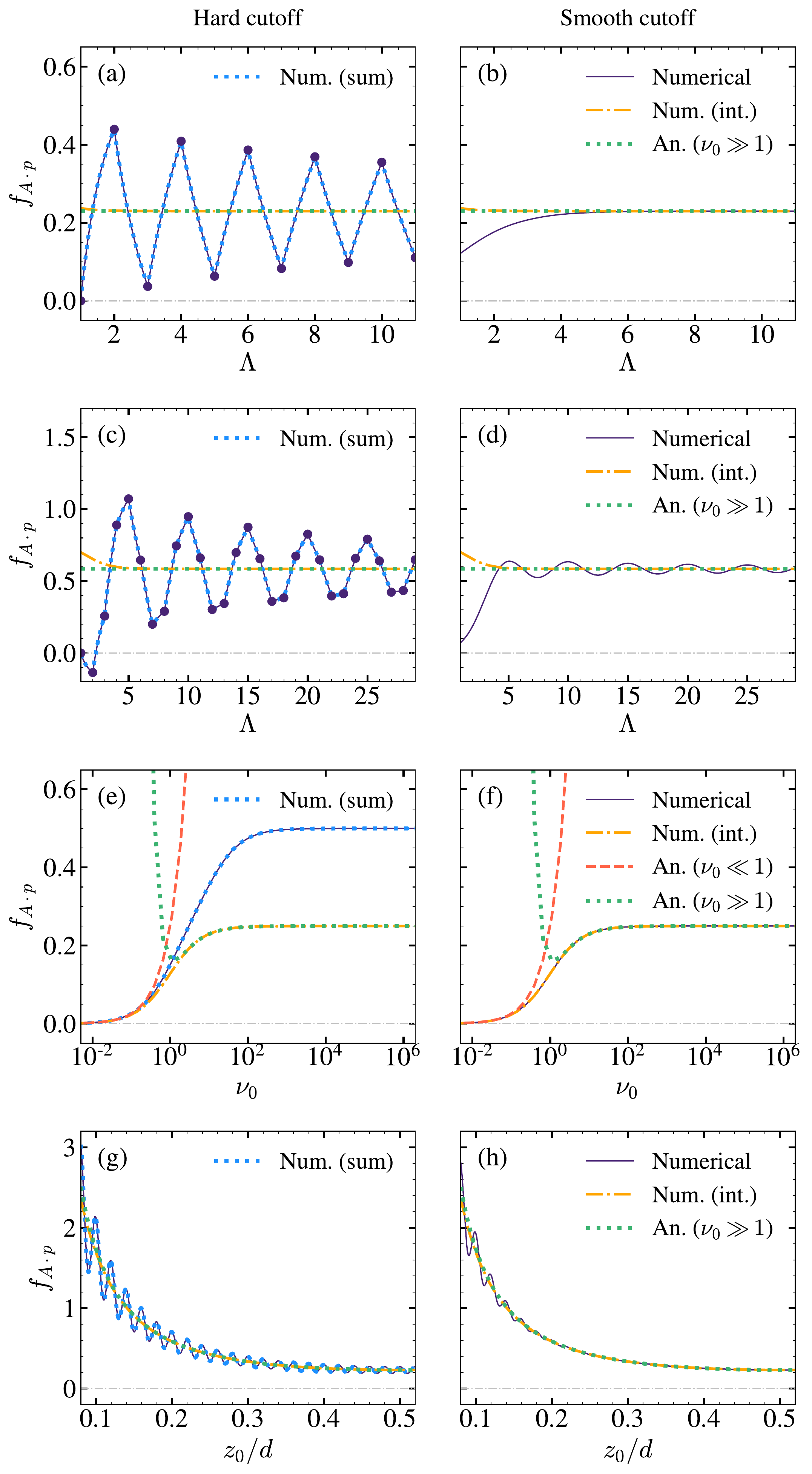}
	\caption{Numerical evaluation of $\funApf$ from \autoref{eq:SMLamb_fmdef} (continuous blue line)
	using either a sharp cutoff at $\nu = \Lambda$ (left panels) or the smooth cutoff function $h_c (\nu/\Lambda) = 1-1/(1+\e^{-\nu+\Lambda})$ (right panels). The dash-dotted yellow line plots \autoref{eq:SMLamb_f0numint}. The contribution $\funApf$ is represented versus $\Lambda$ when $z_0/d = 0.5$ (a-b) and $z_0/d = 0.2$ (c-d) with $\nu_0 = 10$; versus the transition frequency $\nu_0$ when $z_0/d = 0.5$ and $\Lambda = 10$ (e-f); and versus the position of the particle when $\nu_0 = 10$ and $\Lambda = 50$ (g-h). The analytical expressions valid in the limits $\nu_0 \ll 1$ (\autoref{eq:SMLamb_f0anall}) and $\nu_0 \gg 1$ (\autoref{eq:SMLamb_f0anagg}) are depicted with a dashed red and a dotted green lines, respectively.}
	\label{fig:SMLamb_fm}
\end{figure}

To analyze the behavior of $\funApf$ in the low-frequency or strong-confinement limit, $\nu_0\ll1$, we expand the expression in \autoref{eq:SMLamb_f0numsum} to lowest order in $\nu_0$,
\begin{equation} \label{eq:SMLamb_f0anall}
	\funApf (\zeta)  \simeq -\frac{\nu_0}{2} \sum_{n=1}^\infty (-1)^n  \cos (n\zeta) = \frac{\nu_0}{4}.
\end{equation}
This result can be obtained without introducing a cutoff explicitly by using, for example, Borel summation.  In \autoref{fig:SMLamb_fm} we compare this result with the numerical evaluation of $\funApf$ with sharp and smooth cutoff functions. The convergence of this function behaves similar to what we have observed for $\funAAf$, but the oscillations near the walls are less pronounced.

\subsubsection{Summary}
\begin{figure}[th]	
	\centering
	\includegraphics[width=\linewidth]{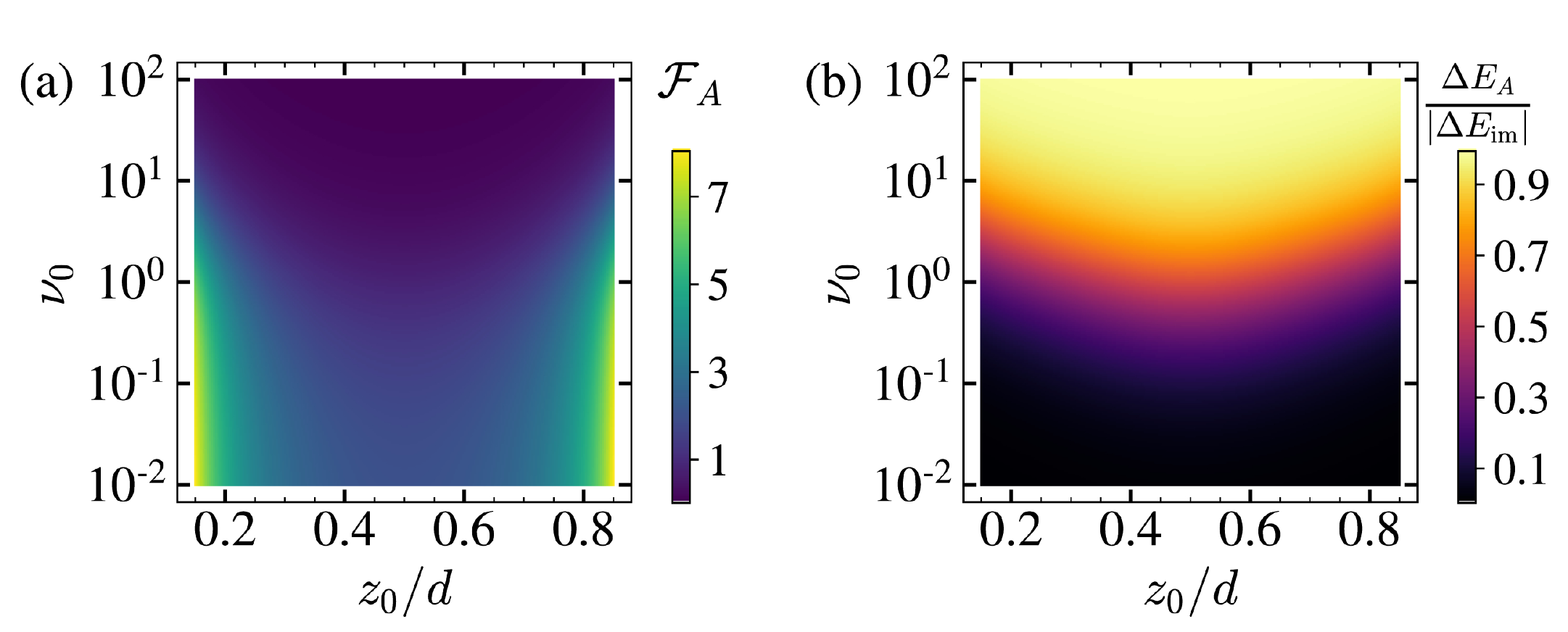}
	\caption{Dimensionless function $\mathcal{F}_{\bA}$ defined in \autoref{eq:SMLamb_prefactorFAdef}  (a) and ratio $\Delta E_{\bA}/|\Delta E_{\rm im}|$ (b) as a function of the position of the dipole, $z_0/d$, and the normalized frequency $\nu_0$.}
	\label{fig:SMLamb_FAFim}
\end{figure}

In summary, we have found cutoff-independent results for all the functions $\funAAg$, $\funAAf$, $\funApg$ and $\funApf$, which appear in the expression for $\Delta E_{\bA}$ in \autoref{eq:SMLamb_DeltaEA_final}. In the limit of strong confinement, $\nu_0\ll1$, we approximately obtain
\begin{equation}
\begin{split}
\mathcal{F}_{\bA}\simeq 2\pi &\left[\frac{1}{12}+ \frac{1}{4\sin^2(\pi z_0/d)}\right.\\
&\left.- \nu_0 \left(\nu_0 + \frac{1}{2}\right) \ln \left(\frac{\nu_0+1}{\nu_0}\right) + \frac{2\nu_0}{3} \right].
\end{split}
\end{equation}
The contributions in the second line, which arise from the $\bp\cdot \bA$ term, vanish as $\nu_0\rightarrow 0$ and therefore we obtain $\mathcal{F}_{\bA}(\nu_0\rightarrow 0,z_0/d=1/2)\approx 2\pi/3$. In the opposite limit, $\nu_0\gg1$, we find instead 
\begin{equation}
\begin{split}
\mathcal{F}_{\bA}\simeq \frac{\mathcal{F}_{\rm im}(z_0/d)}{\pi \nu_0} + o [\nu_0^{-2}].
\end{split}
\end{equation}
Here the first term cancels exactly the electrostatic contribution, $\Delta E_{\rm im}$, such that the total energy shift will be determined by higher-order terms in the expansions in \autoref{eq:SMLamb_gm_highfreq_expand} and \autoref{eq:SMLamb_f0anagg}. This corresponds to the regime of retarded Casimir-Polder interactions, which, in view of the small absolute value of vacuum shifts, is less relevant for the current study. Finally, we can use numerics to evaluate $\mathcal{F}_{\bA}$ also for all intermediate values of $\nu_0$ and the results are summarized in \autoref{fig:SMLamb_FAFim} (a). In \autoref{fig:SMLamb_FAFim} (b), we also show the ratio $\Delta E_{\bA}/|\Delta E_{\rm im}|$, which verifies the bound given in \autoref{eq:Bound} in the main text. 
%

\section{Plasmonic cavities}
\label{sec:SMplasmonicCavities}

In this section we provide a detailed derivation of all the results concerning the plasmonic nanosphere setup discussed in the second part of the main text. Here a single dipole is coupled to the quantized plasmon modes of a metallic nanosphere of radius $R$ and electron density $n_0$. To describe plasmonic excitations, we follow Ref.~\cite{Prodan2004_SM} and model the electron gas as an incompressible, irrotational fluid. Then, for small displacements of this electron fluid, the plasmonic excitations are fully described in terms of the surface charge density $\sigma(t)$. Further, we can assume that for small enough dimensions, interactions between the electrons and between the electrons and the dipole are only due to Coulomb interactions. Therefore, in the following analysis we ignore all interactions with the transverse electromagnetic field, which would contribute additional corrections $\sim O(\alpha)$. 

\subsection{Quantized plasmon modes} 
For an irrotational fluid with $ \nabla \times {\bv}=0$, the velocity field ${\bv}$ can be written as the gradient of a scalar field $\eta$, which obeys the Laplace equation $\nabla^2 \eta=0$ and satisfies the boundary condition~\cite{Prodan2004_SM} 
\begin{equation} \label{eq:sigma_eta}
	\dot \sigma= n_0 e \partial_r \eta.
\end{equation}
In the following we expand the surface density in terms of the spherical harmonics $Y_{\ell,m}$, 
\begin{equation} 
	\sigma(t, \widehat {\bR})= n_0 e  \sum_{\ell=1}^\infty \sum_{m=-\ell}^\ell \sqrt{\frac{4\pi \ell }{3}}   \xi_{\ell,m}(t)  Y_{\ell,m}(\widehat {\bR}).
\end{equation}
Here we introduced the convention $\widehat {\bx}\equiv(\theta,\varphi)$, where $\theta$ and $\varphi$ are the polar and the azimuthal angles of ${\bx}$, respectively. The amplitudes $\xi_{\ell,m}(t)$ are the dynamical degrees of freedom and have units of length. Correspondingly,  for the scalar field we obtain the expansion
\begin{equation}
	\eta({\br},t)  = R \sum_{\ell=1}^\infty \sum_{m=-\ell}^\ell \sqrt{\frac{4\pi}{3  \ell }} \left( \frac{r}{R}\right)^\ell Y_{\ell,m}(\widehat {\br}) \dot \xi_{\ell,m}(t),
\end{equation} 
where we made use of Eq.~\eqref{eq:sigma_eta}.

The Lagrangian of the plasmonic sphere is $\mathcal{L}_P=T_P-V_P$. The kinetic energy is given by~\cite{Prodan2004}
\begin{equation} 
	\begin{split}
		T_P=&\frac{n_0 m_e}{2} \int  \eta \partial_r  \eta \, dS = \frac{M}{2} \sum_{\ell,m} \dot \xi^2_{\ell,m}(t),
	\end{split}
\end{equation} 
where $M= 4\pi R^3n_0m_e/3$ is the total mass of all electrons.  The potential term, $V_P$, accounts for the Coulomb interaction between the surface charges,
\begin{equation}
	V_P= \frac{1}{8\pi \epsilon_0} \int dS \int dS^\prime  \frac{\sigma(\widehat {\br} ) \sigma(\widehat {\br}')}{|{\br} -{\br}'|},
\end{equation} 
where the integration runs over the surface $S$ of the sphere. To evaluate this integral, we make use of the Laplace expansion,
\begin{equation}
	\frac{1}{|{\br} -{\br}'|}  = \frac{4\pi}{R} \sum_{\ell,m}  \frac{1}{2\ell+1}  Y^*_{\ell,m}(\widehat {\br})Y_{\ell,m}(\widehat {\br}'),
\end{equation} 
and obtain 
\begin{equation} 
	\mathcal{L}_P= \frac{M}{2} \sum_{\ell,m}\left[ \dot \xi^2_{\ell,m}(t)- \omega_\ell^2 \xi^2_{\ell,m}(t)\right].
\end{equation} 
Here,
\begin{equation} 
	\omega_\ell = \sqrt{\frac{\ell}{2\ell+1}} \omega_P , \qquad \omega_P=\sqrt{\frac{n_0 e^2}{\epsilon_0 m_e}},
\end{equation} 
are the frequencies of the plamonic eigenmodes of the sphere. Since the plamonic excitations are harmonic, we immediately obtain the 
Hamiltonian of the quantized nanosphere,
\begin{equation}
	H_P= \sum_{\ell,m} \frac{p_{\ell,m}^2}{2M}+ \frac{1}{2}M\omega_\ell^2  \xi^2_{\ell,m} = \sum_{\ell,m}\hbar \omega_{\ell} a^\dag_{\ell,m}a_{\ell,m},
\end{equation} 
where the $p_{\ell,m}$ are the canonical momentum operators and $a_{\ell,m}$ and $a^\dag_{\ell,m}$ are bosonic annihilation and creation operators defined via
\begin{eqnarray}
	\xi_{\ell,m}&=& \sqrt{\frac{\hbar}{2M \omega_\ell}} \left(a^\dag_{\ell,m}+ a_{\ell,m} \right),  \\
	p_{\ell,m}&=&i \sqrt{\frac{\hbar M \omega_\ell}{2}} \left(a^\dag_{\ell,m}- a_{\ell,m} \right).
\end{eqnarray}
The quantized surface charge density is given by
\begin{equation}
	\sigma(t,\widehat {\bR})= n_0 e  \sum_{\ell,m}  \sqrt{\frac{2\pi \hbar \ell}{3M\omega_\ell}} Y_{\ell,m}(\widehat {\bR}) \left( a_{\ell,m}+a^\dag_{\ell,m}\right).
\end{equation} 

\subsection{Dipole-plasmon interactions}
We now consider a point-like dipole with a positive charge $+q$ placed at a fixed distance $D=R+z_0$ away from the center of the sphere. We model the motion of the negative charge as an effective particle moving in a potential $V_d(z_d)$ along the z-direction, but ignore the motion along the $x$ and $y$ directions for simplicity. Then, the total Lagrangian for this system is 
\begin{equation}
	\mathcal{L}=\mathcal{L}_P + \frac{m \dot z_d^2}{2} -V_d(z_d) -V_{\rm int}.
\end{equation}
Here, $V_{\rm int}$ accounts for the electrostatic interaction between the dipole and the surface charges on the sphere, which under the dipole approximation is given by
\begin{equation}
	V_{\rm int}=-\frac{qz_d}{4\pi \epsilon_0} \nabla_D \int dS \frac{\sigma({\br} )}{|D{\bf e}_z- {\br}|}.
\end{equation}
We can use again the Laplace expansion,
\begin{equation}
	\begin{split}
		\frac{1}{|D{\bf e}_z-{\br} |}&= \sum_{\ell,m}  \frac{4\pi}{2\ell+1} \frac{r^\ell}{D^{\ell+1}} Y^*_{\ell,m}(-\widehat{\br})Y_{\ell,m}(0,0)\\
		&=\sum_{\ell}  \sqrt{ \frac{4\pi}{2\ell+1}} \frac{r^\ell}{D^{\ell+1}} Y_{\ell,0}(\widehat{\br}),
	\end{split}
\end{equation}
and write the interaction as
\begin{equation}
	V_{\rm int}=\frac{n_0e qz_d}{4\pi \epsilon_0} \sum_{\ell} \sqrt{ \frac{4\pi}{2\ell+1}} \sqrt{ \frac{4\pi \ell}{3}} (\ell+1) \left( \frac{R}{D} \right)^{\ell+2} \xi_{\ell,0}.
\end{equation}
The total Hamiltonian is then given by 
\begin{equation}\label{eq:PlasmonDipoleHamiltonian}
	H= H_{\rm dip}^0 +  \sum_{\ell=1}^\infty \frac{p_{\ell}^2}{2M}+ \frac{1}{2}M\omega_\ell^2  \xi^2_{\ell} +   \sum_{\ell=1}^\infty  K_{\ell} \xi_{\ell} z_d .
\end{equation} 
where $ H_{\rm dip}^0 $ is the Hamiltonian of the bare dipole and 
\begin{equation}
	K_\ell= \frac{n_0e q}{ \epsilon_0} \sqrt{ \frac{\ell (\ell+1)^2}{6\ell+3}} \left( \frac{R}{D} \right)^{\ell+2}.
\end{equation}
Note that in Eq.~\eqref{eq:PlasmonDipoleHamiltonian} we have already omitted the sum over the uncoupled modes with $m\neq 0$ and redefined $\xi_\ell \equiv \xi_{\ell,0}$ for brevity.

\subsection{Electrostatics}\label{sec:ElectrostaticPlasmons}
From the Hamiltonian in Eq.~\eqref{eq:PlasmonDipoleHamiltonian} we obtain the equation of motion for the dipole,
\begin{equation}\label{eq:EQMDipole}
	\ddot z_d=- \partial_{z_d} V_d(z_d)  -\sum_\ell K_\ell \xi_\ell.
\end{equation}
If we assume that the dipole moves slowly compared to the plasmon frequencies, the plasmon modes follow the dipole adiabatically and assume a value 
\begin{equation}
	\xi_\ell(t) \simeq -\frac{K_\ell}{M\omega_\ell^2}z_d(t).
\end{equation} 
When we plug this result back into~\autoref{eq:EQMDipole}, we can derive an effective potential,
\begin{equation}\label{eq:Vesfull}
	\begin{split}
		V_{\rm im} (z_d) &= - \frac{z_d^2}{2}  \sum_{\ell=1}^\infty \frac{K^2_\ell}{M\omega_\ell^2}\\
		&= -\frac{q^2z_d^2 }{8\pi\epsilon_0 R^3} \sum_{\ell=1}^\infty (\ell+1)^2 \left(\frac{R}{R+z_0} \right)^{2\ell+4}\\
		&=-\frac{q^2z_d^2 }{4\pi\epsilon_0 z_0^3}\mathcal{F}_{\rm im}^{{\rm sp}}(z_0/R),
	\end{split}
\end{equation} 
which just corresponds to the electrostatic potential of a static dipole in front of the sphere. Here we introduced the function
\begin{equation}
	\mathcal{F}_{\rm im}^{{\rm sp}}(x) =\frac{x^3}{2}\sum_{\ell=1}^\infty (\ell+1)^2 \left(\frac{1}{1+x} \right)^{2\ell+4},
\end{equation} 
with an asymptotic value  $\mathcal{F}_{\rm im}^{{\rm sp}}(x\ll1 )\approx 1/8$. Therefore, for $z_0\ll R$, i.e., when the dipole sees an almost flat surface, we recover the usual van der Waals potential,
\begin{equation}
	V_{\rm im}(z_d) \simeq - \frac{q^2z_d^2}{4\pi \epsilon_0 (2z_0)^3}.
\end{equation}
Note that here only the dipole moment along the $z$-direction is taken into account.

\subsection{Cavity QED Hamiltonian}
To derive an appropriate multi-mode cavity QED Hamiltonian for this system we set $z_d = a_0 \mu$ and write the plasmon amplitudes in terms of $a_\ell$ and $a_\ell^\dag$. We obtain
\begin{equation}\label{eq:HPlasmonAllModes} 
	H= H_{\rm dip}^0 + \sum_{\ell=1}^{\infty} \hbar \omega_\ell a_\ell^\dag a_\ell  + \hbar \sum_{\ell=1}^{\infty}  g_\ell (a_\ell+ a_\ell^\dag) \mu.
\end{equation}
Here, we have introduced the coupling constants
\begin{equation}
	g_\ell =  g_P  \frac{\ell+1}{2} \sqrt[4]{\frac{\ell}{2\ell+1}}  \left(\frac{R}{R+z_0} \right)^{\ell+2}, 
\end{equation}
where
\begin{equation}
	\frac{g_P}{\omega_P}= \frac{q a_0}{eR} \sqrt{2\pi \alpha \frac{Z_P}{Z_{\rm vac}}}
\end{equation}
and $ Z_P=1/(\pi\epsilon_0 R \omega_P)$ is the characteristic impedance of the fundamental plasmon mode.

Note that Hamiltonian~\eqref{eq:HPlasmonAllModes} includes only the coupling to dynamical modes, while electrostatic effects arise through virtual excitations of those modes, as illustrated by the adiabatic elimination procedure in \autoref{sec:ElectrostaticPlasmons}. In the following we use this adiabatic elimination for all modes with index $\ell >\ell_{\rm max}$, but we keep the full dynamics of all other modes up to $\ell=\ell_{\rm max}$. We then make use of the general relation~\cite{DeBernardis2018PRA97} 
\begin{equation} 
	-\sum_{\ell=\ell_{\rm max}+1}^\infty \frac{\hbar g_\ell^2}{\omega_\ell} \mu^2 = V_{\rm im} (z_d) +  \sum_{\ell=1}^{\ell_{\rm max}} \frac{\hbar g_\ell^2}{\omega_\ell} \mu^2
\end{equation} 
to rewrite the remaining multi-mode cavity QED Hamiltonian in its canonical form
\begin{equation}
	\begin{split}
		H=& H_{\rm dip}^0+ V_{\rm im} (z_d) + \sum_{\ell=1}^{\ell_{\rm max}} \hbar \omega_\ell a_\ell^\dag a_\ell  \\
		&+ \hbar \sum_{\ell=1}^{\ell_{\rm max}}  g_\ell (a_\ell+ a_\ell^\dag) \mu+  \sum_{\ell=1}^{\ell_{\rm max}}  \frac{\hbar g_\ell^2}{\omega_\ell} \mu^2.
	\end{split} 
\end{equation}
This representation has the advantage that the electrostatic contribution appears explicitly, while the added $P^2$-terms for each dynamical mode avoid an unphysical double counting of static interactions. For a weak and near-resonant coupling to the lowest mode, we can set $\ell_{\rm max}=1$ to obtain the correct single-mode cavity QED Hamiltonian. However, for the current analysis we do not make any restrictions on $\ell_{\rm max}$ yet.

\subsection{Ground state energy}
Since we ignore the corrections from the transverse modes in this calculation, the shift of the ground state, 
\begin{equation}
	\Delta E_{\rm GS}=\Delta E_{\rm im} +\Delta E_P,
\end{equation}
contains a first-order contribution from the image potential, $V_{\rm im}$, and a second-order contribution from the coupling to all plasmon modes up to $\ell_{\rm max}$. The electrostatic contribution can be obtained by averaging the expression in Eq.~\eqref{eq:Vesfull} over the ground-state wavefunction and it reduces to the van der Waals energy for small $z_0$. From the second-order coupling to the plasmon modes we obtain
\begin{equation}
	\Delta E_P= - \sum_{\ell=1}^{\ell_{\rm max}}\sum_{j=1}^\infty  \frac{\hbar g_\ell^2 |\langle 0|\mu|j\rangle|^2 }{\omega_\ell+\omega_j} +  \sum_{\ell=1}^{\ell_{\rm max}}  \frac{\hbar g_\ell^2}{\omega_\ell}  \langle 0|\mu^2 |0\rangle, 
\end{equation} 
where the states $|j\rangle$ are the eigenstates of $H_{\rm dip}^0$ with eigenfrequencies $\omega_j$. To simplify the following analysis we consider a harmonic oscillator with frequency $\omega_0$. Then,
\begin{equation}
	\begin{split}
		\Delta E_P= & \sum_{\ell=1}^{\ell_{\rm max}}  \hbar g_\ell^2 \left[\frac{1}{\omega_\ell}  - \frac{1}{\omega_\ell+\omega_0}\right] = \hbar \omega_0 \sum_{\ell=1}^{\ell_{\rm max}}  \frac{g_\ell^2 }{\omega_\ell(\omega_\ell+\omega_0)}\\
		=&  \alpha \hbar \omega_0 \frac{Z_P}{Z_{\rm vac}}  \left(\frac{qa_0}{eR}\right)^2    \tilde{\mathcal{F}}_P \left(x=\frac{\omega_0}{\omega_P}, y=\frac{z_0}{R}\right),
	\end{split}
\end{equation} 
where
\begin{equation}
	\tilde{\mathcal{F}}_P (x,y)= \frac{\pi}{2}  \sum_{\ell=1}^{\ell_{\rm max}} \frac{(\ell+1)^2  \sqrt{(2\ell+1)/\ell} }{1+x\sqrt{(2\ell+1)/\ell}} \left(\frac{1}{1+y} \right)^{2\ell+4}.
\end{equation}
Importantly, this sum converges as we take the limit $\ell_{\rm max}\rightarrow \infty$, which allows us to obtain a cutoff-free value for the ground state energy shift $\Delta E_{\rm GS}$. Further,
in the low-frequency regime, $\omega_0/\omega_P\ll 1$ and by approximating $\sqrt{(2\ell+1)/\ell}\approx \sqrt{2}$, this sum can be evaluated analytically and scales as 
\begin{equation}
	\lim_{ \ell_{\rm max}\rightarrow \infty}   \tilde{\mathcal{F}}_P(x=0, y\ll 1 )\approx \frac{\pi}{\sqrt{32}y^3}.
\end{equation}
Therefore, we define $\mathcal{F}_P(x,y)= y^3\tilde{\mathcal{F}}_P(x,y)$ and rewrite the energy shift as
\begin{equation}\label{eq:DeltaEPsupp}
	\Delta E_{P}\simeq  \alpha  \hbar \omega_0  \left(\frac{qa_0}{e d}\right)^2  \frac{Z_{\rm eff}}{Z_{\rm vac}}  \mathcal{F}_P \left(\frac{\omega_0}{\omega_P}, \frac{z_0}{R}\right),
\end{equation}
where we introduced the effective impedance 
\begin{equation}
	Z_{\rm eff}= \frac{1}{\pi \epsilon_0 z_0 \omega_P}.
\end{equation}
This result shows that in the full multi-mode calculation, the radius $R$ of the sphere is replaced by the distance $z_0$ as the relevant length scale.

\subsection{Sign of the total energy shift} 
By combining the full results for $\Delta E_{\rm im}$ and for $\Delta E_P$ in \autoref{eq:DeltaEPsupp}, we can write the ratio between dynamical and static corrections as
\begin{equation}
	\frac{\Delta E_P}{|\Delta E_{\rm im}|}=  \frac{\omega_0}{\pi \omega_P} \frac{ \mathcal{F}_P(\omega_0/\omega_P,z_0/R)}{ \mathcal{F}_{\rm im}^P(z_0/R)}<1, 
\end{equation} 
where the upper bound is reached for $\omega_0/\omega_P\gg 1$. This shows that the total energy shift is negative for all parameter regimes.

\end{document}